\documentclass[journal,twoside]{IEEEtran}
\usepackage{cite}

\ifCLASSINFOpdf
\else
\fi
\usepackage[cmex10]{amsmath}
\usepackage{amsthm}
\usepackage{amssymb}
\usepackage[vlined,linesnumbered,ruled]{algorithm2e}

\SetFuncSty{functionfont}
\SetCommentSty{commentfont}
\SetAlFnt{\small}

\ifCLASSOPTIONcompsoc
  \usepackage[caption=false,font=normalsize,labelfont=sf,textfont=sf]{subfig}
\else
  \usepackage[caption=false,font=footnotesize]{subfig}
\fi

\usepackage{threeparttable}

\usepackage{url}

\usepackage{enumerate}
\usepackage{tikz}
\usetikzlibrary{arrows,decorations.pathreplacing}
\usepackage{pgfplots}

\usepackage{multirow}

\usepackage{flushend}

\renewcommand{\mathbf}{\boldsymbol}
\newtheorem{theorem}{Theorem}

\newtheorem{lemma}{Lemma}

\newtheorem{proposition}{Proposition}

\newtheorem*{theorem*}{Theorem}
\newtheorem*{exercise*}{Exercise}
\newtheorem*{corollary*}{Corollary}
\newtheorem*{lemma*}{Lemma}
\newtheorem*{property*}{Property}
\newtheorem*{proposition*}{Proposition}
\newtheorem*{problem*}{Problem}
\newtheorem*{observation*}{Observation}

\theoremstyle{definition}

\theoremstyle{remark}
\newtheorem*{remark}{Remark}

\newcommand{\Arikan}{Ar\i kan}

\newcommand{\RR}{\mathbb{R}}
\newcommand{\Reals}{\mathbb{R}}

\newcommand{\IndexSet}[1]{[\![#1]\!]}

\newcommand{\Prob}[1]{\Pr [#1]}

\DeclareMathOperator{\sgn}{sign}

\newcommand{\abs}[1]{|#1|}

\newcommand{\sfL}{\mathsf{L}}

\newcommand{\sfu}{\mathsf{u}}

\newcommand{\bB}{\mathbf{B}}

\newcommand{\bG}{\mathbf{G}}

\newcommand{\bU}{\mathbf{U}}

\newcommand{\bX}{\mathbf{X}}
\newcommand{\bY}{\mathbf{Y}}

\newcommand{\bb}{\mathbf{b}}

\newcommand{\bm}{\mathbf{m}}

\newcommand{\bu}{\mathbf{u}} 
\newcommand{\bv}{\mathbf{v}} 
\newcommand{\bx}{\mathbf{x}} 
\newcommand{\by}{\mathbf{y}}

\newcommand{\calA}{\mathcal{A}}

\newcommand{\calF}{\mathcal{F}}

\newcommand{\calI}{\mathcal{I}}

\newcommand{\calL}{\mathcal{L}}

\newcommand{\calS}{\mathcal{S}}

\newcommand{\calU}{\mathcal{U}}

\newcommand{\calX}{\mathcal{X}}
\newcommand{\calY}{\mathcal{Y}}

\newcommand{\LR}[2]{ {\Lambda_#1^{(#2)}} }
\newcommand{\LL}[2]{ {\sfL_{#1}^{(#2)}} }
\newcommand{\uu}[2]{ {\sfu_{#1}^{(#2)}} }
\newcommand{\W}[2]{ {W_{#1}^{\left(#2\right)}} }

\newcommand{\PM}[2]{ \mathsf{PM}_{#2}^{(#1)} }

\begin{document}
%
\title{LLR-Based Successive Cancellation List Decoding of Polar Codes}
%
%
%

\author{%
  Alexios~Balatsoukas-Stimming,~\IEEEmembership{Student Member,~IEEE,} 
  Mani~Bastani~Parizi,~\IEEEmembership{Student Member,~IEEE,} 
  and~Andreas~Burg,~\IEEEmembership{Member,~IEEE}%
  \thanks{A. Balatsoukas-Stimming and A. Burg are with the Telecommunications
    Circuits Laboratory (TCL), EPFL. Their research is supported by the Swiss
  National Science Foundation grant 200021\_149447.}%
  \thanks{M. Bastani~Parizi is with the Information Theory Laboratory (LTHI),
    EPFL.  His research is supported by the Swiss National Science Foundation
  grant 200020\_146832.}%
  \thanks{This work has been published in parts in the 39th International
  Conference on Acoustics, Speech and Signal Processing (ICASSP'2014).}%
  \thanks{The authors would like to thank Professor Emre~Telatar, Professor
  Ido~Tal, Jun~Lin, and Bo Yuan for helpful discussions.}
}%

%
%

\markboth{Submitted to IEEE Transactions on Signal Processing in September 2014
-- Revised in March 2015}{Balatsoukas-Stimming \MakeLowercase{\textit{et al.}}:
LLR-Based Successive Cancellation List Decoding of Polar Codes}
%



\maketitle

\begin{abstract}
  We show that successive cancellation list decoding can be formulated
  exclusively using log-likelihood ratios. In addition to numerical stability,
  the log-likelihood ratio based formulation has useful properties which
  simplify the sorting step involved in successive cancellation list decoding.
  We propose a hardware architecture of the successive cancellation list decoder
  in the log-likelihood ratio domain which, compared to a log-likelihood domain
  implementation, requires less irregular and smaller memories. This
  simplification together with the gains in the metric sorter, lead to $56\%$
  to $137\%$ higher throughput per unit area than other recently proposed
  architectures. We then evaluate the empirical performance of the CRC-aided
  successive cancellation list decoder at different list sizes using different
  CRCs and conclude that it is important to adapt the CRC length to the list
  size in order to achieve the best error-rate performance of concatenated polar
  codes.  Finally, we synthesize conventional successive cancellation decoders
  at large block-lengths with the same block-error probability as our proposed
  CRC-aided successive cancellation list decoders to demonstrate that, while our
  decoders have slightly lower throughput and larger area, they have a
  significantly smaller decoding latency.
 \end{abstract}


%
\IEEEpeerreviewmaketitle

\section{Introduction}
%
%
%
%
\IEEEPARstart{I}{n} his seminal work \cite{Arik09}, \Arikan{} constructed the
first class of error correcting codes that can achieve the capacity of
any symmetric binary-input discrete memoryless channel (B-DMC) with efficient
encoding \emph{and decoding} algorithms based on \emph{channel polarization}.
In particular, \Arikan{} proposed a low-complexity successive
cancellation (SC) decoder and proved that the block-error
probability of \emph{polar codes} under SC decoding vanishes as their
block-length increases.  The SC decoder is attractive from an implementation
perspective due to its highly structured nature. Several hardware architectures
for SC decoding of polar codes have recently been presented in the literature
\cite{Leroux11,Raymond13,Pamuk2013,Leroux13,Zhang13,Sarkis14b,Fan14}, the first
SC decoder ASIC was presented in \cite{Mishra2012}, and simplifications of
\Arikan{}'s original SC decoding algorithm are studied in
\cite{Alam11,Zhang12,Sarkis13,Zhang14}. 

Even though the block-error probability of polar codes under SC decoding decays
roughly like $O(2^{-\sqrt{N}})$ as a function of the block-length
$N$ \cite{arikan09rate}, they do not perform well at low-to-moderate
block-lengths.  This is to a certain extent due to the sub-optimality of the SC
decoding algorithm. To partially compensate for this sub-optimality, Tal and
Vardy proposed the successive cancellation list (SCL) decoder whose
computational complexity is shown to scale identically to the SC decoder with
respect to the block-length \cite{Tal11}. 

SCL decoding not only improves the block-error probability of polar codes, but
also enables one to use \emph{modified polar codes} \cite{Tal12,Niu12}
which are constructed by concatenating a polar code with a cyclic redundancy
check (CRC) code as an outer code. Adding the CRC increases neither the
computational complexity of the encoder nor that of the decoder by a notable
amount, while reducing the block-error probability significantly, making the
error-rate performance of the modified polar codes under SCL decoding comparable
to the state-of-the-art LDPC codes \cite{Tal12}. In \cite{Li12} an adaptive
variant of the CRC-aided SCL decoder is proposed in order to further improve the
block-error probability of modified polar codes while maintaining the average
decoding complexity at a moderate level.

The SCL decoding algorithm in \cite{Tal11} is described in terms of likelihoods.
Unfortunately, computations with likelihoods are numerically unstable as they
are prone to underflows. In recent hardware implementations of the SCL
decoder~\cite{Bala14,Lin14,Zhang14b,Yuan14,Lin15} the stability problem was
solved by using log-likelihoods (LLs). However, the use of LLs creates other
important problems, such as an irregular memory with varying number of bits per
word, as well as large processing elements, making these decoders still
inefficient in terms of area and throughput.
\subsection*{Contributions and Paper Outline}
After a background review of polar codes and SCL decoding in
Section~\ref{sec:background}, in Section~\ref{sec:theory} we prove that the SCL
decoding algorithm can be formulated exclusively in the \emph{log-likelihood
ratio} (LLR) domain, thus enabling area-efficient and numerically stable
implementation of SCL decoding. We discuss our SCL decoder hardware architecture
in Section~\ref{sec:sclarch} and leverage some useful properties of
the LLR-based formulation in order to \emph{prune} the radix-$2L$ sorter
(implementing the sorting step of SCL decoding) used in \cite{Bala14,icassp} by
avoiding unnecessary comparisons in Section~\ref{sec:sorting}.
Next, in Section~\ref{sec:results} we see that the LLR-based implementation
leads to a significant reduction of the size of our previous hardware
architecture \cite{Bala14}, as well as to an increase of its maximum operating
frequency. We also compare our decoder with the recent SCL decoder architectures
of \cite{Yuan14,Lin15} and show that our decoder can have more than $100\%$
higher throughput per unit area than those architectures. 

Besides the implementation gains, it is noteworthy that most processing blocks
in practical receivers process the data in the form of LLRs. Therefore, the
LLR-based SCL decoder can readily be incorporated into existing systems while
the LL-based decoders would require extra processing stages to convert the
channel LLRs into LLs.  In fairness, we note that one particular advantage of
LL-based SCL decoders is that the algorithmic simplifications of
\cite{Alam11,Zhang12,Sarkis13,Zhang14} can readily be applied to the SCL decoder
\cite{Sarkis14}, while in order to apply those simplifications to an LLR-based
SCL decoder one has to rely on \emph{approximations} \cite{Lin14b}.

Finally, we show that a CRC-aided SCL decoder can be implemented by
incorporating a CRC unit into our decoder, with almost no additional hardware
cost, in order to achieve significantly lower block-error probabilities. As we
will see, for a fixed information rate, the choice of CRC length is critical in
the design of the modified polar code to be decoded by a CRC-aided SCL decoder.
In Section~\ref{sec:CASCLDresults} we provide simulation results showing that
for small list sizes a short CRC will improve the performance of SCL decoder
while larger CRCs will even degrade its performance compared to a standard polar
code. As the list size gets larger, one can increase the length of CRC in order
to achieve considerably lower block-error probabilities.

An interesting question, which is, to the best of our knowledge, still
unaddressed in the literature, is whether it is better to use SC decoding with
long polar codes or SCL decoding with short polar codes. In
Section~\ref{sec:conclusion} we study two examples of long polar codes that
have the same block-error probability under SC decoding as our $(1024,512)$
modified polar codes under CRC-aided SCL decoding and compare the synthesis
results of the corresponding decoders.
\section{Background} \label{sec:background}
\paragraph*{Notation} Throughout this paper, boldface letters denote vectors.
The elements of a vector $\bx$ are denoted by $x_i$ and $\bx_l^m$ means the
sub-vector $[x_l, x_{l+1}, \dots, x_m]^T$ if $m \ge l$ and the null vector
otherwise. If $\calI = \{i_1,i_2,\dots\}$ is an ordered set of indices,
$\bx_\calI$ denotes the sub-vector $[x_{i_1}, x_{i_2}, \dots ]^T$. For a
positive integer $m$, $\IndexSet{m} \triangleq \{0,1,\cdots,m-1\}$. If $\calS$
is a countable set, $\abs{\calS}$ denotes its cardinality. $\log(\cdot)$ and
$\ln(\cdot)$ denote base-$2$ and natural logarithm respectively. We follow the
standard coding theory notation and denote a code of block-length $N$ and rate
$\frac{K}{N}$ as an ``$(N,K)$ code.''

For $N=2^n$, $n \ge 1$, let $\bU$ be a uniformly distributed random vector in
$\{0,1\}^N$ and suppose the random vector $\bX \in \{0,1\}^N$ is computed from
$\bU$ through the linear transform
\begin{equation}
  \bX = \bG_n \bU, \quad  \text{where} \quad
  \bG_n \triangleq \begin{bmatrix} 1 & 1 \\ 0 & 1 \end{bmatrix}^{\otimes n}
  \bB_n,
  \label{eq:polarEnc}
\end{equation}
where $\otimes n$ denotes the $n$th Kronecker power of the matrix and $\bB_n$ is
the bit-reversal permutation.\footnote{%
  Let $\bv$ and $\bu$ be two length $N=2^n$ vectors and index their elements
  using binary sequences of length $n$, $(b_1,b_2,\dots,b_n) \in \{0,1\}^n$.
  Then $\bv = \bB_n \bu$ iff $v_{(b_1,b_2,\dots,b_n)} =
  u_{(b_n,b_{n-1},\dots,b_1)}$ for $\forall (b_1,b_2,\dots,b_n) \in \{0,1\}^n$.%
}%

If $\bX$ is transmitted via $N$ independent uses of the B-DMC $W: \calX \to
\calY$, where $\calX = \{0,1\}$ is the input alphabet and $W(y|x)$ is the
probability distribution function of the output letter $Y \in \calY$ when $x$ is
transmitted, the conditional distribution of the output vector $\bY \in
\calY^N$ is 
\begin{equation}
  W^N(\by|\bx) \triangleq 
  \Pr[\bY = \by | \bX = \bx] = \prod_{i=0}^{N-1} W(y_i | x_i),
  \label{eq:vectorPhysicalChannel}
\end{equation}
for $\forall \bx \in \calX^N$ and $\by \in \calY^N$.  Equivalently, the
distribution of $\bY$ conditioned on $\{\bU = \bu\}$ is
\begin{equation}
  W_n(\by|\bu) \triangleq \Pr[\bY = \by | \bU = \bu]  = 
  W^N(\by|\bG_n \bu),
  \label{eq:superChannel}
\end{equation}
for $\forall \bu \in \calX^N$ and $\forall \by \in \calY^N$ with
$W^N(\by|\bx)$ as in \eqref{eq:vectorPhysicalChannel}.\footnote{Following the
  convention in probability theory, we denote the realizations of the
  random vectors $\bU$, $\bX$, and $\bY$ as $\bu$, $\bx$, and $\by$
respectively.}

`Synthesize' $N$ B-DMCs, $\W{n}{i}, i \in \IndexSet{N}$ by defining $\W{n}{i}$
as the B-DMC whose input is $U_i$ and whose output is the vector of physical
channel outputs $\bY$ together with all preceding elements of $\bU$,
$\bU_0^{i-1}$ as side information, considering all following elements of $\bU$
as i.i.d. Bernoulli noise. Thus, the transition probabilities of $\W{n}{i}:
\calX \to \calY \times \calX^i$ are
\begin{equation}
  \W{n}{i}(\by, \bu_0^{i-1} | u_i) \triangleq \sum_{\bu_{i+1}^{N-1}\in
  \calX^{N-i-1}} \frac{1}{2^{N-1}} W_n(\by|\bu).
  \label{eq:synthChannelDef}
\end{equation}

\Arikan{} shows that as $n \to \infty$, these synthetic channels \emph{polarize}
to `easy-to-use' B-DMCs \cite[Theorem~1]{Arik09}. That is, all except a
vanishing fraction of them will be either almost-noiseless channels (whose
output is almost a deterministic function of the input) or useless channels
(whose output is almost statistically independent of the input). Furthermore,
the fraction of almost-noiseless channels is equal to the symmetric capacity of
the underlying channel---the highest rate at which reliable communication is
possible through $W$ when the input letters $\{0,1\}$ are used with equal
frequency \cite{shannon48}.
\subsection{Polar Codes and Successive Cancellation Decoding} 
Having transformed $N$ identical copies of a `moderate' B-DMC $W$ into $N$
`extremal' B-DMCs $\W{n}{i}, i \in \IndexSet{N}$, \Arikan{} constructs
capacity-achieving \emph{polar codes} by exploiting the almost-noiseless
channels to communicate information bits.
\subsubsection{Polar Coding}
In order to construct a polar code of rate $R$ and block length $N$ for a
channel $W$, the indices of the $NR$ least noisy synthetic channels $\W{n}{i}, i
\in \IndexSet{N}$  are selected as the \emph{information} indices denoted by
$\calA \subset \IndexSet{N}$. The sub-vector $\bu_{\calA}$ will be set to the
$NR$ data bits to be sent to the receiver and $\bu_{\calF}$, where $\calF =
\IndexSet{N} \setminus \calA$, is fixed to some \emph{frozen} vector which is
known to the receiver. The vector $\bu$ is then encoded to the codeword $\bx$
through \eqref{eq:polarEnc} using $O(N \log N)$ binary additions (cf.
\cite[Section~VII]{Arik09}) and transmitted via $N$ independent uses of the
channel $W$. 

The receiver observes the channel output vector $\by$ and estimates the elements
of the $\bu_\calA$ \emph{successively} as follows: Suppose the
information indices are ordered as $\calA= \{i_1, i_2,\dots, i_{NR}\}$ (where
$i_j < i_{j+1}$). Having the channel output, the receiver has all the required
information to decode the input of the synthetic channel $\W{n}{i_1}$
as $\hat{u}_{i_1}$, as, in particular, $\bu_{0}^{i_1-1}$ is a part of the known
sub-vector $\bu_{\calF}$. Since this synthetic channel is assumed to be
almost-noiseless by construction, $\hat{u}_{i_1} = u_{i_1}$ with high
probability. Subsequently, the decoder can proceed to index $i_2$ as the
information required for decoding the input of $\W{n}{i_2}$ is now available.
Once again, this estimation is with high probability error-free. As  detailed
in Algorithm~\ref{alg:sc}, this process is continued until all the information
bits have been estimated.
\begin{algorithm}[htb]
   \For{$i = 0,1,\dots,N-1$}{%
   \eIf(\tcp*[f]{frozen bits}){$i \not\in \calA$}{%
      $\hat{u}_i \gets u_i$\;
    }(\tcp*[f]{information bits}){%
      $\hat{u}_i \gets \arg\max_{u_i \in \{0,1\}} \W{n}{i}(\by,
      \hat{\bu}_0^{i-1} | u_i)$\; \nllabel{lin:sc:decide} 
    }
  }
  \Return $\hat{\bu}_\calA$ \;
  \caption{SC Decoding \cite{Arik09}.}
  \label{alg:sc}
\end{algorithm}
\subsubsection{SC Decoding as a Greedy Tree Search Algorithm} \label{sec:SCvsML}
Let 
\begin{equation}
  \calU(\bu_\calF) \triangleq \{\bv \in \calX^N: \bv_\calF = \bu_\calF\}
  \label{eq:searchSet}
\end{equation}
denote the set of $2^{NR}$ possible length-$N$ vectors that the transmitter can
send.  The elements of $\calU(\bu_\calF)$ are in one-to-one correspondence
with $2^{NR}$ leaves of a binary tree of height $N$: the leaves are constrained
to be reached from the root by following the direction $u_i$ at all levels $i
\in \calF$.  Therefore, any decoding procedure is essentially equivalent to
picking a \emph{path} from the root to one of these leaves on the binary tree. 

In particular, an optimal ML decoder, associates each path with its likelihood
(or any other \emph{path metric} which is a monotone function of the likelihood)
and picks the path that maximizes this metric by exploring \emph{all} possible
paths:
\begin{equation}
  \textstyle
  \hat{\bu}_{\rm ML} = \arg\max_{\bv \in \calU(\bu_\calF)} W_n(\by|\bv).
  \label{eq:mldec}
\end{equation}
Clearly such an optimization problem is computationally infeasible as the number
of paths, $|\calU(\bu_\calF)|$, grows exponentially with the block-length
$N$.

The SC decoder, in contrast, finds a sub-optimal solution by maximizing the
likelihood via a \emph{greedy} one-time-pass through the tree: starting from the
root, at each level $i \in \calA$, the decoder extends the existing path by
picking the child that maximizes the \emph{partial likelihood}
$\W{n}{i}(\by,\hat{\bu}_0^{i-1}| u_i)$.  
\subsubsection{Decoding Complexity} \label{sec:scimpl}
The computational task of the SC decoder is to calculate the pairs of
likelihoods $\W{n}{i}(\by,\hat{\bu}_0^{i-1}|u_i),~u_i \in \{0,1\}$ needed for
the decisions in line~\ref{lin:sc:decide} of Algorithm~\ref{alg:sc}. Since the
decisions are binary, it is sufficient to compute the \emph{decision
log-likelihood ratios (LLRs)},
\begin{equation} 
  \LL{n}{i} \triangleq \ln \biggl( \frac{\W{n}{i}(\by, \hat{\bu}_{0}^{i-1}|0)}
  {\W{n}{i}(\by, \hat{\bu}_{0}^{i-1}|1)} \biggr), \qquad i \in \IndexSet{N}.
  \label{eq:llrDef}
\end{equation}

It can be shown (see \cite[Section~VII]{Arik09} and \cite{Leroux11}) that  the
decision LLRs \eqref{eq:llrDef} can be computed via the recursions,
\begin{align*}
  \LL{s}{2i} & = f_-\big( \LL{s-1}{2i - [i \bmod 2^{s-1}]},
  \LL{s-1}{2^{s} + 2 i - [i \bmod 2^{s-1}]} \big), \\
  \LL{s}{2i+1} & = f_+\big( \LL{s-1}{2i - [i \bmod 2^{s-1}]},
  \LL{s-1}{2^{s} + 2 i - [i \bmod 2^{s-1}]}, \uu{s}{2i} \big),
\end{align*}
for $s = n, n-1, \dots,1$, where $f_-: \Reals^2 \to \Reals$ and $f_+: \Reals^2
\times \{0,1\} \to \Reals$ are defined as
\begin{subequations}
  \begin{align}
    f_-(\alpha,\beta) & \triangleq \ln \Bigl(\frac{e^{\alpha+\beta} +
    1}{e^\alpha + e^\beta}\Bigr), \label{eq:fminus} \\
    f_+(\alpha,\beta,u) & \triangleq (-1)^u \alpha + \beta,
  \end{align}
\end{subequations}
respectively. The recursions terminate at $s=0$ where
\begin{equation*} 
  \LL{0}{i} \triangleq \ln \Bigl(\frac{W(y_i|0)}{W(y_i|1)}\Bigr), \qquad
  \forall i \in \IndexSet{N},
\end{equation*}
are \emph{channel LLRs}. The \emph{partial sums} $\uu{s}{i}$ are computed
starting from $\uu{n}{i} \triangleq \hat{u}_i, ~ \forall i \in
\IndexSet{N}$
and setting
\begin{align*}
  \uu{s-1}{2i - [i \bmod 2^{s-1}]} &= \uu{s}{2i} \oplus \uu{s}{2i+1}, \\
  \uu{s-1}{2^s + 2i - [i \bmod 2^{s-1}]} &=  \uu{s}{2i+1},
\end{align*}
for $s = n, n-1, \dots, 1$.

\begin{figure}[htb]
  \centering
  \scalebox{1}{\begin{tikzpicture}[font=\footnotesize,
    x=1cm,y=-0.75cm,
    >=stealth',
  block/.style={rounded corners,draw=black,fill=blue!20},
  fplus/.style={thick,<-,dashed,orange!80!black},
  fminus/.style={thick,<-,dotted,blue} ]
  \node[block,fill=black!20] at (6,0) (l00) {$\LL{0}{0}$};
  \node[block,fill=black!20] at (6,1) (l01) {$\LL{0}{1}$};
  \node[block,fill=black!20] at (6,2) (l02) {$\LL{0}{2}$};
  \node[block,fill=black!20] at (6,3) (l03) {$\LL{0}{3}$};
  \node[block,fill=black!20] at (6,4) (l04) {$\LL{0}{4}$};
  \node[block,fill=black!20] at (6,5) (l05) {$\LL{0}{5}$};
  \node[block,fill=black!20] at (6,6) (l06) {$\LL{0}{6}$};
  \node[block,fill=black!20] at (6,7) (l07) {$\LL{0}{7}$};
 
  \node	at (6,8) {$s = 0$};

  \node[block] at (4,0) (l10) {$\LL{1}{0}$};
  \node[block] at (4,1) (l11) {$\LL{1}{1}$};
  \node[block] at (4,2) (l12) {$\LL{1}{2}$};
  \node[block] at (4,3) (l13) {$\LL{1}{3}$};
  \node[block] at (4,4) (l14) {$\LL{1}{4}$};
  \node[block] at (4,5) (l15) {$\LL{1}{5}$};
  \node[block] at (4,6) (l16) {$\LL{1}{6}$};
  \node[block] at (4,7) (l17) {$\LL{1}{7}$};
  
  \node	at (4,8) {$s = 1$};
  
  \node[block] at (2,0) (l20) {$\LL{2}{0}$};
  \node[block] at (2,1) (l21) {$\LL{2}{1}$};
  \node[block] at (2,2) (l22) {$\LL{2}{2}$};
  \node[block] at (2,3) (l23) {$\LL{2}{3}$};
  \node[block] at (2,4) (l24) {$\LL{2}{4}$};
  \node[block] at (2,5) (l25) {$\LL{2}{5}$};
  \node[block] at (2,6) (l26) {$\LL{2}{6}$};
  \node[block] at (2,7) (l27) {$\LL{2}{7}$};

  \node	at (2,8) {$s = 2$};
  
  \node[block,fill=green!20] at (0,0) (l30) {$\LL{3}{0}$};
  \node[block,fill=green!20] at (0,1) (l31) {$\LL{3}{1}$};
  \node[block,fill=green!20] at (0,2) (l32) {$\LL{3}{2}$};
  \node[block,fill=green!20] at (0,3) (l33) {$\LL{3}{3}$};
  \node[block,fill=green!20] at (0,4) (l34) {$\LL{3}{4}$};
  \node[block,fill=green!20] at (0,5) (l35) {$\LL{3}{5}$};
  \node[block,fill=green!20] at (0,6) (l36) {$\LL{3}{6}$};
  \node[block,fill=green!20] at (0,7) (l37) {$\LL{3}{7}$};

  \node	at (0,8) {$s = 3$};
  
  \draw[dotted,darkgray] (1,8) -- (1,-0.5);
  \draw[dotted,darkgray] (3,8) -- (3,-0.5);
  \draw[dotted,darkgray] (5,8) -- (5,-0.5);
  
  \draw[fminus] (l30.east) -- (l20.west);
  \draw[fminus] (l30.east) -- (l24.west);
  \draw[fminus] (l20.east) -- (l10.west);
  \draw[fminus] (l20.east) -- (l12.west);
  \draw[fminus] (l24.east) -- (l14.west);
  \draw[fminus] (l24.east) -- (l16.west);
  \draw[fminus] (l10.east) -- (l00.west);
  \draw[fminus] (l10.east) -- (l01.west);
  \draw[fminus] (l12.east) -- (l02.west);
  \draw[fminus] (l12.east) -- (l03.west);
  \draw[fminus] (l14.east) -- (l04.west);
  \draw[fminus] (l14.east) -- (l05.west);
  \draw[fminus] (l16.east) -- (l06.west);
  \draw[fminus] (l16.east) -- (l07.west);
  \draw[fplus] (l31.east) -- (l20.west);
  \draw[fplus] (l31.east) -- (l24.west);
  \draw[fminus] (l32.east) -- (l21.west);
  \draw[fminus] (l32.east) -- (l25.west);
  \draw[fplus] (l21.east) -- (l10.west);
  \draw[fplus] (l21.east) -- (l12.west);
  \draw[fplus] (l25.east) -- (l14.west);
  \draw[fplus] (l25.east) -- (l16.west);
  \draw[fplus] (l33.east) -- (l21.west);
  \draw[fplus] (l33.east) -- (l25.west);
  \draw[fminus] (l34.east) -- (l22.west);
  \draw[fminus] (l34.east) -- (l26.west);
  \draw[fminus] (l22.east) -- (l11.west);
  \draw[fminus] (l22.east) -- (l13.west);
  \draw[fminus] (l26.east) -- (l15.west);
  \draw[fminus] (l26.east) -- (l17.west);
  \draw[fplus] (l11.east) -- (l00.west);
  \draw[fplus] (l11.east) -- (l01.west);
  \draw[fplus] (l13.east) -- (l02.west);
  \draw[fplus] (l13.east) -- (l03.west);
  \draw[fplus] (l15.east) -- (l04.west);
  \draw[fplus] (l15.east) -- (l05.west);
  \draw[fplus] (l17.east) -- (l06.west);
  \draw[fplus] (l17.east) -- (l07.west);
  \draw[fplus] (l35.east) -- (l22.west);
  \draw[fplus] (l35.east) -- (l26.west);
  \draw[fminus] (l36.east) -- (l23.west);
  \draw[fminus] (l36.east) -- (l27.west);
  \draw[fplus] (l23.east) -- (l11.west);
  \draw[fplus] (l23.east) -- (l13.west);
  \draw[fplus] (l27.east) -- (l15.west);
  \draw[fplus] (l27.east) -- (l17.west);
  \draw[fplus] (l37.east) -- (l23.west);
  \draw[fplus] (l37.east) -- (l27.west);

  \draw[decorate,decoration={brace,amplitude=10pt}] (6.5,0) -- (6.5,7)
  node[black,midway,rotate=270,yshift=2em]{Channel LLRs -- stage $s=0$};
  
  \draw[decorate,decoration={brace,amplitude=10pt,mirror}] (-0.5,0) -- (-0.5,7)
  node[black,midway,rotate=90,yshift=2em]{Decision LLRs -- stage $s=n$};

\end{tikzpicture}}
  \caption{The butterfly computational structure of the SC decoder for $n=3$;
    blue and orange arrows show $f_-$ and $f_+$ updates respectively.}
  \label{fig:scbutterfly}
\end{figure}
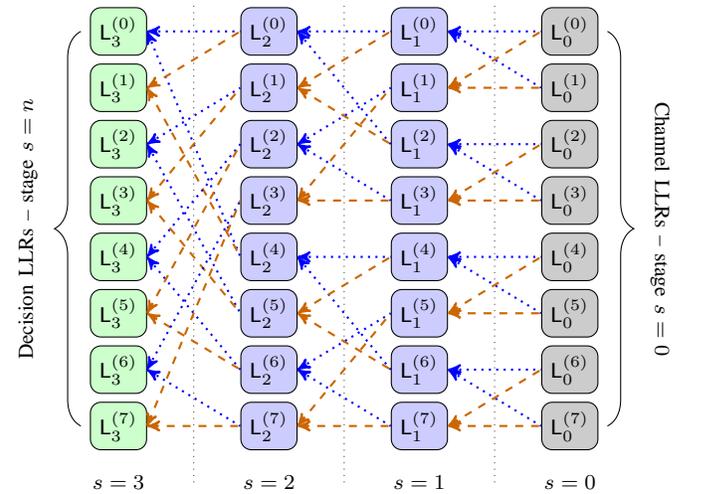

Therefore, the entire set of $N \log N$ LLRs $\LL{s}{i}, s \in \{1,\dots,n\}, i
\in \IndexSet{N}$ can be computed using $O(N \log N)$ \emph{update}s since from
each pair of LLRs at \emph{stage} $s$, a pair of LLRs at stage $s+1$ is
calculated using $f_-$ and $f_+$ update rules (see
Figure~\ref{fig:scbutterfly}).  Additionally the decoder must keep track of $N
\log N$ partial sums $\uu{s}{i},s \in \IndexSet{n}, i \in \IndexSet{N}$  and
update them after decoding each bit $\hat{u}_i$. 

\begin{remark}
  It can easily be checked that (cf. \cite{Leroux11})
  \begin{equation}
    f_-(\alpha, \beta) \approx \tilde{f}_-(\alpha, \beta) \triangleq
    \sgn(\alpha) \sgn(\beta) \min \{\abs{\alpha}, \abs{\beta}\},
    \label{eq:fminustilde}
  \end{equation}
  where $\tilde{f}_-$ is a `hardware-friendly' function as it involves only the
  easy-to-implement $\min\{\cdot,\cdot\}$ operation (compared to $f_-$ which
  involves exponentiations and logarithms). For a hardware implementation of the
  SC decoder the update rule $f_-$ is replaced by $\tilde{f}_-$. Given  $f_+$,
  such an approximation is called the ``min-sum approximation'' of the decoder.
\end{remark}

\subsection{Successive Cancellation List Decoding} \label{sec:listSCIntro}
The \emph{successive cancellation list (SCL)} decoding algorithm, introduced in
\cite{Tal11}, converts the greedy one-time-pass search of SC decoding into a
breadth-first search under a complexity constraint in the following way: At each
level $i \in \calA$, instead of extending the path in only one direction, the
decoder is duplicated in two parallel \emph{decoding threads} continuing in
either possible direction. However, in order to avoid the exponential growth of
the number of decoding threads, as soon as the number of parallel decoding
threads reaches $L$, at each step $i \in \calA$, only $L$ threads corresponding
the $L$ most likely paths (out of $2L$ tentatives) are
retained.\footnote{Although it is not necessary, $L$ is normally a power of
$2$} The decoder eventually finishes with a \emph{list} of $L$ candidates
$\hat{\bu}[\ell],~\ell\in\IndexSet{L}$, corresponding to $L$ (out of $2^{NR}$)
paths on the binary tree and declares the most likely of them as the final
estimate. This procedure is formalized in Algorithm~\ref{alg:scl}. Simulation
results in \cite{Tal11} show that for a $(2048,1024)$ polar code, a relatively
small list size of $L=32$ is sufficient to have a close-to-ML block-error
probability.
 
\newcommand{\AT}{\calL}
\begin{algorithm}[htb]
  \SetKwFunction{duplicate}{duplicatePath}
  $\AT \gets \{0\}$ \tcp*[r]{start with a single active thread}
  \For{$i = 0,1,\dots,N-1$}{%
    \eIf(\tcp*[f]{frozen bits})
    {$i \not\in \calA$}{%
      $\hat{u}_i[\ell] \gets u_i$ for $\forall \ell \in \AT$\;
    }(\tcp*[f]{information bits}){%
      \eIf(\tcp*[f]{duplicate all the threads}){$|\AT| < L$}{%
	\ForEach {$\ell \in \AT$}
	{%
	  \duplicate{$\ell$}\;  \nllabel{lin:scl:dup1}
	}
      }{%
	Compute $P_{\ell,u} = \W{n}{i}(\by, \hat{\bu}_0^{i-1}[\ell]|u)$, for
	$\forall \ell \in \AT$ and $\forall u \in\{0,1\}$\;
	\nllabel{lin:scl:decideStart} \nllabel{lin:scl:metric}
	$\tau \gets \text{the median of $2L$ numbers $P_{\ell,u}$}$\;
	\nllabel{lin:scl:median}
	\ForEach {$\ell \in \AT$ such that $P_{\ell,0} < \tau$ and
	$P_{\ell,1} < \tau$}{%
	  Kill the thread $\ell$ and set $\AT \gets \AT \setminus
	  \{\ell\}$\;
	}
	\For{$\ell \in \AT$}{%
	  \uIf{$P_{\ell,u} > \tau$ while $P_{\ell, u \oplus 1} < \tau$}{%
	    $\hat{u}_i[\ell] \gets u$\;
	  }
	  \Else(\tcp*[f]{both $P_{\ell,0}$ and $P_{\ell,1}$ are $\ge \tau$})
	  {%
	    \duplicate{$\ell$}\;
	    \nllabel{lin:scl:decideEnd} \nllabel{lin:scl:dup2}
	  }
	}
      }
    }
  }
  $\ell^* \gets \arg\max_{\ell \in \AT} \W{n}{N-1}(\by,\hat{\bu}_0^{N-1}[\ell]|
  \hat{u}_N[\ell])$\; \nllabel{lin:scl:chooseML}
  \Return $\hat{\bu}_\calA[\ell^*]$\;
  \vspace*{0.5ex}
  \SetKwProg{subroutine}{subroutine}{}{}
  \subroutine{\duplicate{$\ell$}}{
    Copy the thread $\ell$ into a new thread $\ell' \not\in \AT$\;
    $\AT \gets \AT \cup \{\ell'\}$\;
    $\hat{u}_i[\ell] \gets 0$\;
    $\hat{u}_i[\ell'] \gets 1$\;
  }
  \caption{SC List Decoding \cite{Tal11}}
  \label{alg:scl}
\end{algorithm}
While a naive implementation of SCL decoder would have a decoding complexity of
at least $\Omega(L \cdot N^2)$ (due to $\Theta(L \cdot N)$ \emph{duplications}
of data structures of size $\Omega(N)$ in lines~\ref{lin:scl:dup1} and
\ref{lin:scl:dup2} of Algorithm~\ref{alg:scl}), a clever choice of data
structures together with the recursive nature of computations enables the
authors of \cite{Tal11} to use a copy-on-write mechanism and implement the
decoder in $O(L \cdot N \log N)$ complexity.
\subsection{CRC-Aided Successive Cancellation List Decoder}
\label{sec:CRCAidedSCLD}
In an extended version of their work \cite{Tal12},  Tal and Vardy observe that
when the SCL decoder fails, in most of the cases, the correct path
(corresponding to $\bu_{\calA}$) is among the $L$ paths the decoder has
ended up with. The decoding error happens since there exists another more likely
path which is selected in line~\ref{lin:scl:chooseML} of
Algorithm~\ref{alg:scl} (note that in such situations the ML decoder would also
fail). They, hence, conclude that the performance of polar codes would be
significantly improved if the decoder were assisted for its final choice.

Such an assistance can be realized by adding $r$ more non-frozen bits (i.e.,
creating a polar code of rate $R+r/N$ instead of rate $R$) to the
underlying polar code and then setting the last $r$ non-frozen bits to an
$r$-bit CRC  of the first $NR$ information bits (note that the \emph{effective}
information rate of the code is unchanged). The SCL decoder, at
line~\ref{lin:scl:chooseML}, first discards the paths that do not pass
the CRC and then chooses the most likely path among the remaining ones.  Since
the CRC can be computed efficiently \cite[Chapter~7]{macwilliams1978theory},
this does not notably increase the computational complexity of the decoder. The
empirical results of \cite{Tal12} show that a $(2048,1024)$ concatenated polar
code (with a $16$-bit CRC) decoded using a list decoder with list size
of $L=32$, outperforms the existing state-of-the-art WiMAX $(2304,1152)$ LDPC
code \cite{wimax}.

\begin{remark} According to \cite{Tal14}, the empirical results of \cite{Tal12}
  on the CRC-aided successive cancellation list decoder (CA-SCLD) are obtained
  using a $(2048,1040)$ (outer) polar code with the last $16$ unfrozen bits
  being the CRC of the first $1024$ information bits and the results on the
  non-CRC aided (standard) SCL decoder are obtained using a $(2048,1024)$ polar
  code---both having an effective information rate of $\frac12$. In
  \cite{Niu12,Lin14,Lin15} the CA-SCLD is realized by keeping the number of
  non-frozen bits fixed and setting the last $r$ of them to the CRC of the
  preceding $NR-r$ information bits. This reduces the effective information rate
  of the code and makes the comparison between the SCLD and the
  CA-SCLD unfair.\footnote{In
    \cite{Li12} this discrepancy is not clarified. However, this work focuses
    only on CA-SCLD without comparison of the performance of a SCLD to a
  CA-SCLD.} 
\end{remark}

\section{LLR-Based Path Metric Computation} \label{sec:theory}
Algorithms~\ref{alg:sc} and \ref{alg:scl} are both valid high-level descriptions
of SC and SCL decoding, respectively.  However, for implementing these
algorithms, the stability of the computations is crucial.  Both algorithms
summarized in Section~\ref{sec:background} are described in terms of likelihoods
which are \emph{not} safe quantities to work with; a decoder implemented using
the likelihoods is prone to underflow errors as they are typically tiny
numbers.\footnote{%
  As noticed  in \cite{Tal12}, it is not difficult to see that
  $\W{n}{i}(\by, \bu_0^{i-1}|u_i) \le 2^{-i}$.%
}
 
Considering the binary tree picture that we provided in
Section~\ref{sec:SCvsML}, the decision LLRs $\LL{n}{i}$ \eqref{eq:llrDef}
summarize all the necessary information for choosing the most likely child among
two children of the same parent at level $i$. In Section~\ref{sec:scimpl} we saw
that having this type of decisions in the conventional SC decoder allows us to
implement the computations in the LLR domain using numerically stable
operations.  However, in the SCL decoder, the problem is to choose the $L$ most
likely children out of $2L$ children of $L$ different parents
(lines~\ref{lin:scl:decideStart} to \ref{lin:scl:decideEnd} of
Algorithm~\ref{alg:scl}). For these comparisons the decision log-likelihood
\emph{ratios} $\LL{n}{i}$ alone are not sufficient. 

Consequently, the software implementation of the decoder in \cite{Tal11}
implements the decoder in the likelihood domain by rewriting the recursions of
Section~\ref{sec:scimpl} for computing pairs of likelihoods
$\W{n}{i}(\by, \hat{\bu}_0^{i-1}|u_i)$, $u_i \in \{0,1\}$ from pairs of
channel likelihoods $W(y_i|x_i), x_i \in \{0,1\}, i \in \IndexSet{N}$. To avoid
underflows, at each intermediate step of the updates the likelihoods are scaled
by a common factor such that $P_{\ell,u}$ in line~\ref{lin:scl:metric} of
Algorithm~\ref{alg:scl} is proportional to $W(\by, \hat{\bu}_0^{i-1}[\ell] | u)$
\cite{Tal12}.

Alternatively, such a normalization step can be avoided by performing the
computations in the log-likelihood (LL) domain, i.e., by computing the pairs
$\ln \big(\W{n}{i}(\by, \hat{\bu}^{i-1}[\ell]|u)\big)$, $u \in \{0,1\}$ for
$i\in \IndexSet{N}$ as a function of channel log-likelihood pairs
$\ln(W(y_i|x_i))$, $x_i \in \{0,1\}$, $i \in \IndexSet{N}$ \cite{Bala14}.
Log-likelihoods provide some numerical stability, but still involve some issues
compared to the log-likelihood \emph{ratios} as we shall discuss in
Section~\ref{sec:sclarch}.

Luckily, we shall see that the decoding paths can still be ordered according to
their likelihoods using all of the past decision LLRs $\LL{n}{j}$,
$j\in \{0,1\cdots,i\}$ and the trajectory of each path as summarized in the
following theorem. 
\begin{theorem} \label{thm:metric}
  For each path $\ell$ and each level $i \in \IndexSet{N}$ let the
  \emph{path-metric} be defined as:
  \begin{equation} \label{eq:pathMetricDef}
    \PM{i}{\ell} \triangleq \sum_{j = 0}^{i} \ln \bigl( 1 + e^{-(1-2
      \hat{u}_j[\ell]) \cdot \LL{n}{j}[\ell] } \bigr),
  \end{equation}
  where 
  \begin{equation*}
    \LL{n}{i}[\ell] = \ln \bigg( \frac{\W{n}{i}(\by, \hat{\bu}^{i-1}[\ell]|0)}
    {\W{n}{i}(\by, \hat{\bu}^{i-1}[\ell]|1)} \bigg),
  \end{equation*}
  is the log-likelihood ratio of bit $u_i$ given the channel output $\by$ and
  the past trajectory of the path $\hat{\bu}_0^{i-1}[\ell]$.
  
  If all the information bits are uniformly distributed in $\{0,1\}$, for
  any pair of paths $\ell_1, \ell_2$,
  $$\W{n}{i}(\by,\hat{\bu}^{i-1}[\ell_1]|\hat{u}_i[\ell_1]) <
  \W{n}{i}(\by,\hat{\bu}^{i-1}[\ell_2]|\hat{u}_i[\ell_2])$$ if and only if
  $$\PM{i}{\ell_1} > \PM{i}{\ell_2}.$$
\end{theorem}
In view of Theorem~\ref{thm:metric}, one can implement the SCL decoder using $L$
parallel low-complexity \emph{and stable} LLR-based SC decoders as the
underlying building blocks and, in addition, keep track of $L$ path-metrics. The
metrics can be updated successively as the decoder proceeds by setting
\begin{subequations}\label{eq:pmupdate}
\begin{equation}
  \PM{i}{\ell} = \phi\big(\PM{i-1}{\ell}, \LL{n}{i}[\ell],
  \hat{u}_i[\ell]\bigr),
\end{equation}
where the function $\phi: \RR_+^2 \times \{0,1\} \to \RR_+$ is defined as
\begin{equation}
  \phi(\mu, \lambda, u) \triangleq \mu + \ln \bigl( 1 + e^{-(1-2 u) \lambda}
  \bigr). \label{eq:pmupdatefunc} 
\end{equation}
\end{subequations}
As shown in Algorithm~\ref{alg:sclLLR}, the paths can be compared based on their
likelihood using the values of the associated path metrics.
\begin{algorithm}[htb]
  \SetKwFunction{duplicate}{duplicatePath}
  $\AT \gets \{0\}$ \tcp*[r]{start with a single active thread} 
  $\PM{0}{0} \gets 0$ \;
  \For{$i = 0,1,\dots,N-1$}{%
    Compute $\LL{n}{i}[\ell]$ for $\forall \ell \in \AT$ \tcp*[l]{parallel SC
    decoders} \nllabel{lin:sclLLR:LLR}
    \eIf(\tcp*[f]{frozen bits})
    {$i \not\in \calA$}{%
      $\bigl(\hat{u}_i[\ell], \PM{i}{\ell}\bigr) \gets \bigr(u_i,
      \phi(\PM{i-1}{\ell},\LL{n}{i}[\ell],u_i) \bigl)$ for $\forall \ell \in
      \AT$ 
      \tcp*[r]{cf. \eqref{eq:pmupdatefunc}}
      \nllabel{lin:sclLLR:frozenUpdate}
    }(\tcp*[f]{information bits}){%
      Set $P_{\ell,u} \gets \phi(\PM{i-1}{\ell}, \LL{n}{i},u)$ for $\forall \ell
      \in \AT$ and $\forall u \in\{0,1\}$ \tcp*[r]{cf \eqref{eq:pmupdatefunc}}
      \nllabel{lin:sclLLR:P}
      \eIf(\tcp*[f]{duplicate all the threads}){$|\AT| < L$}{%
	\ForEach {$\ell \in \AT$}
	{%
	  \duplicate{$\ell$}\; 
	}
      }{%
	$\tau \gets \text{the median of $2L$ numbers $P_{\ell,u}$}$\;
	\nllabel{lin:sclLLR:median}
	\ForEach {$\ell \in \AT$ such that $P_{\ell,0} > \tau$ and
	$P_{\ell,1} > \tau$}{%
	  Kill the thread $\ell$ and set $\AT \gets \AT \setminus
	  \{\ell\}$\;
	}
	\For{$\ell \in \AT$}{%
	  \uIf{$P_{\ell,u} > \tau$ while $P_{\ell, u \oplus 1} < \tau$}{%
	    $\bigl(\hat{u}_i[\ell], \PM{i}{\ell}\bigr) \gets (u,P_{\ell,u})$\;
	  }
	  \Else(\tcp*[f]{both $P_{\ell,0}$ and $P_{\ell,1}$ are $\le \tau$})
	  {%
	    \duplicate{$\ell$}\;
	  }
	}
      }
    }
  }
  $\ell^* \gets \arg\min_{\ell \in \AT} \PM{N}{\ell}$\;
  \Return $\hat{\bu}_\calA[\ell^*]$\;
  \vspace*{0.5ex}
  \SetKwProg{subroutine}{subroutine}{}{}
  \subroutine{\duplicate{$\ell$}}{
    Copy the thread $\ell$ into a new thread $\ell' \not\in \AT$\;
    $\AT \gets \AT \cup \{\ell'\}$\;
    \nllabel{lin:sclLLR:copy}
    $\bigl(\hat{u}_i[\ell], \PM{i}{\ell}\bigr) \gets (0, P_{\ell,0})$\;
    \nllabel{lin:sclLLR:ext0}
    $\bigl(\hat{u}_i[\ell'], \PM{i}{\ell'}\bigr) \gets (1, P_{\ell,1})$\;
    \nllabel{lin:sclLLR:ext1}
  }
  \caption{LLR-based formulation of SCL Decoding}
  \label{alg:sclLLR}
\end{algorithm}

Before proving Theorem~\ref{thm:metric} let us provide an intuitive
interpretation of our metric. Since 
\begin{equation*}
  \ln(1+e^x) \approx \begin{cases} 0 & \text{if } x < 0, \\ 
    x & \text{if } x \ge 0,
  \end{cases}
\end{equation*}
the update rule \eqref{eq:pmupdate} is well-approximated if we replace $\phi$
with $\tilde\phi: \RR_+^2 \times \{0,1\} \to \RR_+$ defined as 
\begin{equation} 
  \tilde\phi(\mu, \lambda, u) \triangleq \begin{cases}
    \mu & \text{if $u = \frac12[1-\sgn(\lambda)]$,} \\
    \mu + \abs{\lambda} & \text{otherwise.}
  \end{cases}
  \label{eq:pmupdateApprox}
\end{equation}
We also note that $\frac12[1-\sgn(\LL{n}{i}[\ell])]$ is the direction that the
LLR (given the past trajectory $\hat{\bu}_0^{i-1}[\ell]$) suggests. This is the
same decision that a SC decoder would have taken if it was to estimate the
value of $u_i$ at step $i$ given the past set of decisions
$\hat{\bu}_0^{i-1}[\ell]$ (cf. line~\ref{lin:sc:decide} in
Algorithm~\ref{alg:sc}).  Equation~\eqref{eq:pmupdateApprox} shows that if at
step $i$ the $\ell$th path does not follow the direction suggested by
$\LL{n}{i}[\ell]$ it will be penalized by an amount $\approx |\LL{n}{i}[\ell]|$.

Having such an interpretation, one might immediately conclude that the path that
SC decoder would follow will always have the lowest penalty hence is always
declared as the output of the SCL decoder. So why should the SCL
decoder exhibit a better performance compared to the SC decoder? The answer is
that such a reasoning is correct only if \emph{all} the elements of $\bu$ are
information bits. As soon as the decoder encounters a frozen bit, the path
metric is updated based on the likelihood of that frozen bit, given the past
trajectory of the path and the a-priori known value of that bit (cf.
line~\ref{lin:sclLLR:frozenUpdate} in Algorithm~\ref{alg:sclLLR}). This can
penalize the SC path by a considerable amount, if the value of that frozen bit
does not agree with the LLR given the past trajectory (which is an
indication of a preceding erroneous decision), while keeping some other paths
unpenalized.

We devote the rest of this section to the proof of Theorem~\ref{thm:metric}.
\begin{lemma} 
  \label{lem:prop}
  If $U_i$ is uniformly distributed in $\{0,1\}$, then, 
  $$ \frac{ \W{n}{i}(\by, \bu_0^{i-1}|u_i)}
  {\Prob{\bU_0^{i} = \bu_0^{i}| \bY = \by}}  = 2\Prob{\bY=\by}.$$ 
\end{lemma}
\begin{IEEEproof}
  Since $\Prob{U_i = u_i} = \frac12$ for $\forall u_i \in \{0,1\}$,
  \begin{align*}
    & \frac{\W{n}{i}(\by, {\bu}_0^{i-1} | u_i)}{\Prob{\bU_0^i = \bu_0^i| \bY =
    \by}} = \frac{\Prob{\bY = \by, \bU_0^i= \bu_0^i}} 
    {\Prob{U_i=u_i} \Prob{\bU_0^i = \bu_0^i | \bY=\by}} \\
    & \quad = \frac{\Prob{\bY=\by}\Prob{\bU_0^{i} = \bu_0^{i} |
    \bY=\by}}{\Prob{U_i = u_i} \Prob{\bU_0^i=\bu_0^i | \bY=\by}} =  2
    \Prob{\bY=\by}.
    \tag*{\IEEEQEDhere}
  \end{align*} 
\end{IEEEproof}
\begin{IEEEproof}[Proof of Theorem~\ref{thm:metric}]
  It is sufficient to show
  \begin{equation}
    \PM{i}{\ell} = -\ln\left( \Prob{\bU_0^{i} = \hat{\bu}_0^i[\ell] | \bY = \by}
    \right). 
    \label{eq:pathMetric}
  \end{equation}
  Having shown \eqref{eq:pathMetric}, Theorem~\ref{thm:metric} will follow as
  an immediate corollary to Lemma~\ref{lem:prop} (since the channel output $\by$
  is fixed for all decoding paths). Since the path index $\ell$ is
  fixed on both sides of \eqref{eq:pathMetricDef} we
  will drop it in the sequel.  Let 
  \begin{equation*}
    \LR{n}{i} \triangleq \frac{\W{n}{i}(\by,\hat{\bu}_0^{i-1}|0)}
    {\W{n}{i}(\by,\bu_0^{i-1}|1)} =
    \frac{\Prob{\bY=\by,\bU_0^{i-1}=\hat{\bu}_0^{i-1}, U_i=0}} {\Prob{\bY=\by,
    \bU_0^{i-1}=\hat{\bu}_0^{i-1}, U_i=1}}
  \end{equation*}
  (the last equality follows since $\Prob{U_i=0} = \Prob{U_i=1}$),  and observe
  that showing \eqref{eq:pathMetric} is equivalent to proving
  \begin{equation}
    \Prob{\bU^i = \hat{\bu}^i | \bY = \by} = \prod_{j=0}^{i} \bigl(
    1 + (\LR{n}{j})^{-(1-2\hat{u}_j)} \bigr)^{-1}.
    \label{eq:expmetric}
  \end{equation}
  Since
  \begin{align*} 
    & \Prob{\bY=\by, \bU_0^{i-1}=\hat{\bu}_0^{i-1}} 
    = \sum_{\hat{u}_i \in \{0,1\}} \Prob{\bY=\by, \bU_0^{i}=\hat{\bu}_0^{i}}
    \\
    & \quad = \Prob{\bY=\by, \bU_0^{i}=\hat{\bu}_0^{i}} 
    \bigl( 1 + (\LR{n}{i})^{-(1-2\hat{u}_i)} \bigr),
  \end{align*}
  \begin{align}      
    & \Prob{\bY=\by, \bU_0^i = \hat{\bu}_0^i}  \nonumber \\
    & \, = \bigl(1 + (\LR{n}{i})^{-(1-2\hat{u}_i)}\bigr)^{-1}
    \Prob{ \bY=\by, \bU_0^{i-1}=\hat{\bu}_0^{i-1} }.     
    \label{eq:onestep}
  \end{align}
  Repeated application of \eqref{eq:onestep} (for $i-1, i-2, \dots, 0$)
  yields
  \begin{equation*}      
    \Prob{\bY=\by, \bU_0^i = \hat{\bu}_0^i} 
    = \prod_{j=0}^i \bigl(1 + (\LR{n}{j})^{ -(1-2\hat{u}_i)} \bigr)^{-1}
    \Prob{ \bY=\by}.
  \end{equation*}
  Dividing both sides by $\Prob{\bY=\by}$ proves
  \eqref{eq:expmetric}.
\end{IEEEproof}
\begin{figure*} 
  \centering 
  \begin{minipage}[t]{0.68\textwidth} 
    \includegraphics[width=1\textwidth]{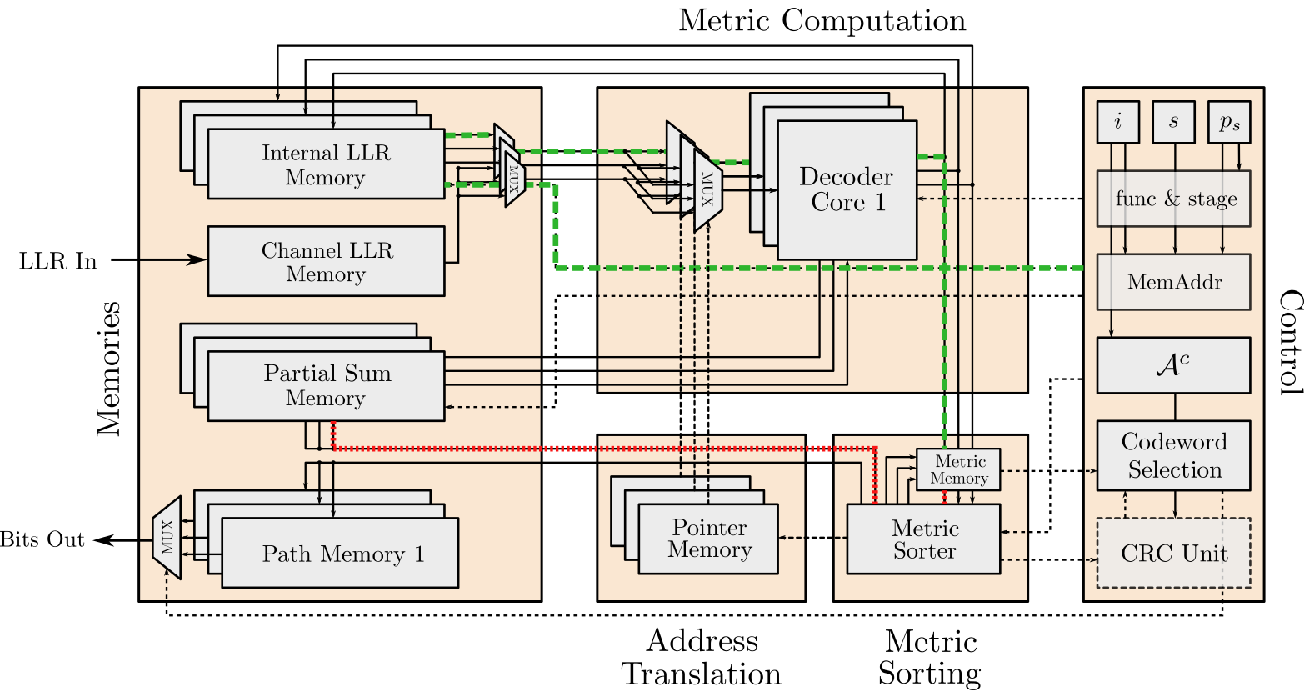} 
    \caption{Overview of the SCL decoder architecture. Details on the $i,s,p_s,$
      as well as the func \& stage and MemAddr components inside the control
      unit, which are not described in this paper, can be found in
      \cite{Bala14}. The dashed green and the dotted red line show the
      critical paths for $L=2$ and $L=4,8$ respectively.
    }
    \label{fig:scldecarch}
	\end{minipage}
	$\quad$
	\begin{minipage}[t]{0.28\textwidth}
  \includegraphics[width=1\textwidth]{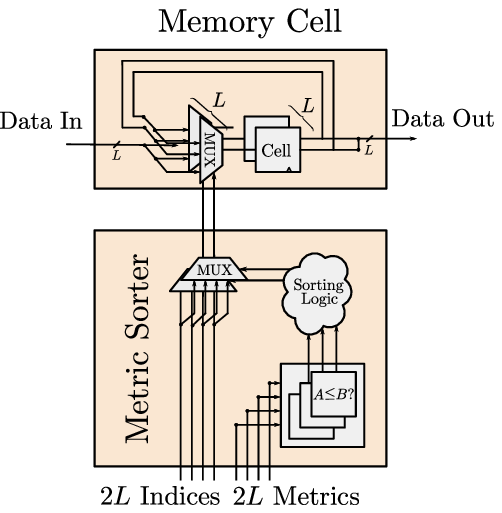}
	\caption{Bit-cell copying mechanism controlled by the metric sorter.}\label{fig:cellcopying}
	\end{minipage}
\end{figure*}
\section{SCL Decoder Hardware Architecture}\label{sec:sclarch}
In this section, we show how the LLR-based path metric which we derived in 
the previous section can be exploited in order to derive a very efficient 
LLR-based SCL decoder hardware architecture. More specifically, we give a 
detailed description of each unit of our LLR-based SCL decoder architecture,
which essentially consists of $L$ parallel SC decoders along with a path
management unit which coordinates the tree search. Moreover, we highlight the
advantages over our previous LL-based architecture described in~\cite{Bala14}. 
Our SCL decoder consists of five units: the \emph{memories unit}, the
\emph{metric computation unit} (MCU), the \emph{metric sorting unit}, the
\emph{address translation unit}, and the \emph{control unit}. An overview of the
SCL decoder is shown in Figure~\ref{fig:scldecarch}.
\subsection{LLR and Path Metric Quantization}\label{subsec:quantization}
All LLRs are quantized using a $Q$-bit signed uniform quantizer with step size
$\Delta = 1$. The path metrics are unsigned numbers which are quantized using
$M$ bits. Since the path metrics are initialized to $0$ and, in the worst case,
they are incremented by $2^{Q-1}-1$ for each bit index $i$, the maximum possible
value of a path metric is $N (2^{Q-1}-1) = 2^{n + Q - 1} - 2^n < 2^{n + Q - 1}$.
Hence, at most $M = n + Q - 1$ bits are sufficient to ensure that there will be
no overflows in the path metric. In practice, any path that gets continuously
harshly penalized will most likely be discarded. Therefore, as we will see in
Section~\ref{sec:results}, much fewer bits are sufficient in practice for the
quantization of the path metrics.
\subsection{Metric Computation Unit}
The computation of the LLRs  (line \ref{lin:sclLLR:LLR} of
Algorithm~\ref{alg:sclLLR}) can be fully parallelized.  Consequently, the MCU
consists of $L$ parallel SC decoder cores which implement the SC decoding update
rules and compute the $L$ decision LLRs using the semi-parallel SC decoder 
architecture of~\cite{Leroux13} with $P$ processing elements (PEs). 
These decision LLRs are required to update the path
metrics $\PM{i}{\ell}$. Whenever the $L$ decision LLRs have been computed, the
MCUs wait for one clock cycle. During this single clock cycle, the path metrics
$\PM{i}{\ell}$ are updated and sorted. Moreover, based on the result of metric
sorting, the partial sum, path, and pointer memories are also updated in the
same clock cycle, as described in the sequel.

Each decoder core reads its input LLRs from one of the $L$ physical LLR memory
banks based on an address translation performed by the pointer memory (described
in more detail in Section~\ref{sec:pointermem}).
\subsection{Memory Unit}
\subsubsection{LLR Memory}\label{subsubsec:llrmem}
The channel LLRs are fixed during the decoding process of a given codeword,
meaning that an SCL decoder requires only one copy of the channel LLRs. These
are stored in a memory which is $\frac{N}{P}$ words deep and $QP$ bits wide.
On the other hand, the internal LLRs of the intermediate stages of the SC
decoding (metric computation) process are different for each path $\ell \in
\IndexSet{L}$. Hence we require $L$ physical LLR memory banks with $N-1$ memory
positions per bank. All LLR memories have two reads ports, so that all $P$ PEs
can read their two $Q$-bit input LLRs simultaneously. 
Here, register based storage cells are used to implement all the memories.
\subsubsection{Path Memory}
The path memory consists of $L$ $N$-bit registers, denoted by
$\hat{\bu}[\ell],~\ell \in \IndexSet{L}$. When a path $\ell$ needs to be
duplicated, the contents of $\hat{\bu}[\ell]$ are copied to $\hat{\bu}[\ell']$,
where $\ell'$ corresponds to an inactive path (cf.  line~\ref{lin:sclLLR:copy}
of Algorithm~\ref{alg:sclLLR}). The decoder is stalled for one clock cycle in
order to perform the required copy operations by means of $N$ $L \times L$
crossbars which connect each $\hat{\bu}[\ell],~\ell \in \IndexSet{L}$ with all
other $\hat{\bu}[\ell'],~\ell' \in \IndexSet{L}$. The copy mechanism is
presented in detail in Figure~3, where we show how each memory bit-cell is
controlled based on the results of the metric sorter. After path $\ell$ has been
duplicated, one copy is extended with the bit value $\hat{u}_i[\ell] = 0$, while
the other is updated with $\hat{u}_i[\ell'] = 1$ (cf.
lines~\ref{lin:sclLLR:ext0} and \ref{lin:sclLLR:ext1} of
Algorithm~\ref{alg:sclLLR}).
\subsubsection{Partial Sum Memory}
The partial sum memory consists of $L$ PSNs, where each PSN is implemented as
in \cite{Leroux13}. When a path $\ell \in \IndexSet{L}$ needs to be duplicated,
the contents of the PSN $\ell$ are copied to another PSN $\ell'$, where $\ell'$
corresponds to an inactive path (cf.  line~\ref{lin:sclLLR:copy} of
Algorithm~\ref{alg:sclLLR}). Copying is performed in parallel with the copy of
the path memory in a single clock cycle by using $N$ $L \times L$ crossbars
which connect each PSN $\ell \in \IndexSet{L}$ with all other PSNs $\ell' \in
\IndexSet{L}$.  If PSN $\ell$ was duplicated, one copy is updated with the bit
value $\hat{u}_i[\ell] = 0$, while the other copy is updated with
$\hat{u}_i[\ell'] = 1$. If a single copy of PSN $\ell$ was kept, then this copy
is updated with the value of $\hat{u}_i[\ell]$ that corresponds to the surviving
path.
\subsection{Address Translation Unit}\label{sec:pointermem}
The copy-on-write mechanism used in \cite{Tal11} (which is fully
applicable to LLRs) is sufficient to ensure that the decoding complexity is $O(L
N \log N)$, but it is not ideal for a hardware implementation as, due to the
recursive implementation of the computations, it still requires copying the
internal LLRs which is costly in terms of power, decoding latency, and silicon
area. On the other hand, a sequential implementation of the computations
enables a more hardware-friendly solution~\cite{Bala14}, where each path has its
own virtual internal LLR memory, the contents of which are physically spread
across all of the $L$ LLR memory banks. The translation from virtual memory to
physical memory is done using a small~\emph{pointer memory}. 
When a path $\ell$ needs to be duplicated, as with the partial sum memory, the
contents of row $\ell$ of the pointer memory are copied to some row
corresponding to a discarded path through the use of $L \times L$ crossbars.
\subsection{Metric Sorting Unit}
The metric sorting unit contains a \emph{path metric memory} and a \emph{path
metric sorter}. The path metric memory stores the $L$ path metrics
$\PM{i}{\ell}$ using $M$ bits of quantization for each metric. In order to find
the median $\tau$ at each bit index $i$ (line \ref{lin:sclLLR:median} of
Algorithm~\ref{alg:sclLLR}), the path metric sorter sorts the $2L$ candidate
path metrics $P_{\ell,u}$, $\ell \in \IndexSet{L}$, $u \in \{0,1\}$
(line~\ref{lin:sclLLR:P} of Algorithm~\ref{alg:sclLLR}). The path
metric sorter takes the $2L$ path metrics as an input and produces the sorted
path metrics, as well as the path indices $\ell$ and bit values $u$ which
correspond to the sorted path metrics as an output. Since decoding can not
continue before the surviving paths have been selected, the metric sorter is a
crucial component of the SCL decoder. Hence, we will discuss the sorter
architecture in more detail in Section~\ref{sec:sorting}.
\subsection{Control Unit}
The control unit generates all memory read and write addresses as in
\cite{Leroux13}.  Moreover, the control unit contains the codeword selection
unit and the optional CRC unit.

The CRC unit contains $L$ $r$-bit CRC memories, where $r$ is the number of CRC
bits. A bit-serial implementation of a CRC computation unit is very efficient in
terms of area and path delay, but it requires a large number of clock cycles to
produce the checksum. However, this computation delay is masked by the
bit-serial nature of the SCL decoder itself and, thus, has no impact on the
number of clock cycles required to decode each codeword. Before decoding each
codeword, all CRC memories are initialized to $r$-bit all-zero vectors. For
each $\hat{u}_i[\ell],~i \in \calA$, the CRC unit is activated to update the
CRC values. When decoding finishes, the CRC unit declares which
paths $\ell \in \IndexSet{L}$ pass the CRC. When a path is duplicated the 
corresponding CRC memory is copied by means of $L\times L$ crossbars 
(like the partial sums and the path memory).

If the CRC unit is present, the codeword selection unit selects the most likely
path (i.e., the path with the lowest metric) out of the paths that pass the
CRC. Otherwise, the codeword selection unit simply chooses the most likely path.
\subsection{Clock Cycles Per Codeword}
Let the total number of cycles required for metric sorting at all information
indices $i \in \calA$ be denoted by $D_{\mathrm{MS}}(\calA)$. As we will see
in Section~\ref{sec:sortingLatency}, the sorting latency depends on the number
of information bits and may depend on the pattern of frozen and information
bits as well (both of these parameters can be deduced given $\calA$). Then, our
SCL decoder requires 
\begin{equation} 
  D_{\mathrm{SCL}}(N,P,\calA) = 2N + \frac{N}{P}\log \frac{N}{4P} +
  D_{\mathrm{MS}}(\calA) \label{eq:latency}
\end{equation}
cycles to decode each codeword.
\subsection{Advantages Over LL-based SCL Decoder Implementation}
The LLs in the SCL decoders of~\cite{Bala14,Lin14,Zhang14b,Yuan14,Lin15} are all
positive numbers and the corresponding LL-domain update rules involve only
additions and comparisons.  This means that, as decoding progresses through the
decoding stages, the dynamic range of the LLs is increased. Thus, in order to
avoid catastrophic overflows, all LLs in stage $s$ are quantized using $Q+s$
bits. In the LLR-based implementation of this paper, the LLRs of all stages can
be quantized using the same number of bits since the update rules involve both
addition and subtraction and the dynamic range of the LLRs in different stages
is smaller than that of the LLs.  This leads to a regular memory where all
elements have the same bit-width.  Hence, as we will see in
Section~\ref{sec:results}, using LLRs significantly reduces the total size of
the decoder. In addition, the PEs in the LL-based SCL decoder architectures of
\cite{Bala14,Lin14} must support computations with a much larger bit-width than
the ones in our LLR-based SCL decoder architecture.  Moreover, it turns out that
the path metric in the LLR-based decoder can be quantized using much fewer bits
than in the LL-based decoder, hence decreasing the delay and the size of the
comparators in the metric sorting unit. Finally, the LLR-based formulation
enables us to significantly simplify the metric sorter, as explained in the
following section.
\section{Simplified Sorter}\label{sec:sorting}
For large list sizes ($L \ge 4$), the maximum (critical) delay path passes
through the metric sorter, thus reducing the maximum operating frequency of the
decoder in \cite{Bala14,icassp}.  It turns out that the LLR-based path metric we
introduced in Theorem~\ref{thm:metric} has some properties (which the LL-based
path metric lacks) that can be used to simplify the sorting task. 

To this end, we note that the $2L$ real numbers that have to be sorted in
line~\ref{lin:sclLLR:median} of Algorithm~\ref{alg:sclLLR} are not arbitrary;
half of them are the previously existing path-metrics (which can be assumed to
be already sorted as a result of decoding the preceding information bit) and the
rest are obtained by adding positive real values (the absolute value of the
corresponding LLRs) to the existing path metrics. Moreover, we do not need to
sort \emph{all} these $2L$ potential path metrics; a sorted list of the $L$
smallest path metrics is sufficient.

Hence, the sorting task of the SCL decoder can be formalized as follows: Given a
sorted list of $L$ numbers 
\begin{equation*}
  \mu_0 \le \mu_1 \le \cdots \le \mu_{L-1}
\end{equation*}
a list of size $2L$, $\bm = [m_0,m_1,\cdots,m_{2L-1}]$ is created by setting 
\begin{equation*}
  m_{2 \ell }  := \mu_\ell \quad \text{and} \quad
  m_{2 \ell + 1}  := \mu_\ell + a_\ell, 
  \qquad \ell \in \IndexSet{L},
\end{equation*}
where $a_\ell \ge 0$, for $\forall \ell \in \IndexSet{L}$.  The problem is to
find a sorted list of $L$ smallest elements of $\bm$ when the elements of $\bm$
have the following two properties: for $\forall \ell \in \{0,1,\cdots,L-2\}$,
\begin{subequations}%
  \begin{align}%
    m_{2 \ell} & \le m_{2 (\ell + 1)}, 
    \label{eq:oldListSorted} \\
    m_{2 \ell} & \le m_{2 \ell + 1}. 
    \label{eq:positiveIncrements}%
  \end{align}%
\end{subequations}%

\subsection{Full Radix-$2L$ Sorter}
\begin{figure}[tb]
  \centering
  \subfloat[Full Radix-$2L$ Sorter]{%
    \includegraphics[scale=0.96]{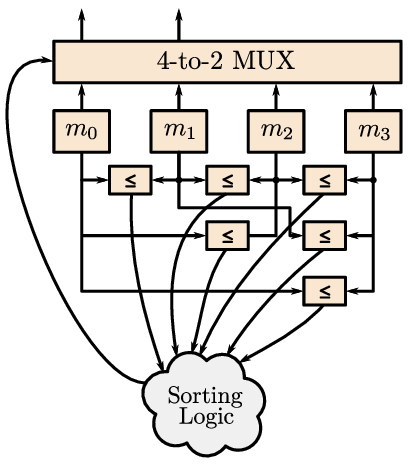}%
    \label{fig:fullRadixSorter}
  }\quad
  \subfloat[Pruned Radix-$2L$ Sorter]{%
    \includegraphics[scale=0.96]{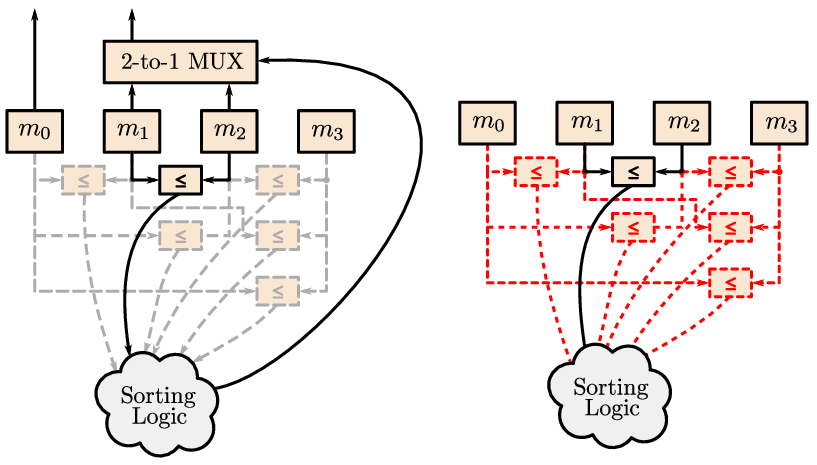}
    \label{fig:prunedRadixSorter}
  }
  \caption{Radix-$2L$ sorters for $L=2$}
\end{figure}
The most straightforward way to solve our problem is to sort the list $\bm$ up
to the $L$-th element. This can be done using a simple extension of the 
radix-$2L$ sorter described in~\cite{Amaru12}, which blindly compares
every pair of elements $(m_\ell, m_{\ell'})$ and then combines the results to
find the first $L$ smallest elements. This is the solution we used in
\cite{Bala14}, which requires $\binom{2L}{2} = L(2L-1)$ comparators together
with $L$ $2L$-to-$1$ multiplexers (see Figure~\ref{fig:fullRadixSorter}). The 
\emph{sorting logic} combines the results of all comparators in order to 
generate the control signal for the  multiplexers (cf. \cite{Amaru12} for
details).  The maximum path delay of the radix-$2L$ sorter is mainly determined
by the complexity of the sorting logic, which in turn depends on the number of
comparator results that need to be processed.

\subsection{Pruned Radix-$2L$ Sorter} \label{sec:prunedRadix2L}
The \emph{pruned radix-$2L$} sorter presented in this section reduces the 
complexity of the sorting logic of the radix-$2L$ sorter and, thus, also the 
maximum path delay, by eliminating some pairwise comparisons whose results 
are either already known or irrelevant.

\begin{proposition} \label{prop:prunedRadix}
  It is sufficient to use a pruned radix-$2L$ sorter that involves only
  $(L-1)^2$ comparators to find the $L$ smallest elements of $\bm$. This sorter
  is obtained by 
  \begin{enumerate}[(a)]
    \item removing the comparisons between every even-indexed element of $\bm$
      and all following elements, and 
    \item removing the comparisons between $m_{2L-1}$ and all other elements of
      $\bm$.
  \end{enumerate}
\end{proposition}
\begin{IEEEproof}
  Properties \eqref{eq:oldListSorted} and \eqref{eq:positiveIncrements} imply 
  $m_{2\ell} \le m_{\ell'}$ for $\forall \ell' > 2\ell$.  Hence, the outputs of
  these comparators are known. Furthermore, as we only need the first $L$
  elements of the list sorted and $m_{2L-1}$ is never among the $L$ smallest
  elements of $\bm$, we can always replace $m_{2L-1}$ by $+\infty$ (pretending
  the result of the comparisons involving $m_{2L-1}$ is known) without affecting
  the output of the sorter.

  In step (a) we have removed $\sum_{\ell=0}^{L-1} (2L-1 - 2\ell) = L^2$
  comparators and in step (b) $(L-1)$ comparators (note that in the full sorter
  $m_{2L-1}$ is compared to all $(2L-1)$ preceding elements but $L$ of them
  correspond to even-indexed elements whose corresponding comparators have
  already been removed in step (a)). Hence we have $L (2L-1) - L^2 - (L-1) =
  (L-1)^2$ comparators. 
\end{IEEEproof}

Besides the $(L-1)^2$ comparators, the pruned radix-$2L$ sorter requires $L-1$
$(2L-2)$-to-$1$ multiplexers (see Figure~\ref{fig:prunedRadixSorter}).  

The pruned radix-$2L$ sorter is derived based on the assumption that the
existing path metrics are already sorted.  This assumption is violated when the
decoder reaches the first frozen bit after the first cluster of information
bits; at each frozen index, some of the path-metrics are unchanged and some are
increased by an amount equal to the absolute value of the LLR. 
In order for the assumption to hold when the decoder reaches the next cluster of
information bits, the $L$ existing path metrics have to be sorted before the
decoding of this cluster starts.  The existing pruned radix-$2L$ sorter can be
used for sorting $L$ arbitrary positive numbers as follows.
\begin{proposition} 
  Let $a_0, a_1, \dots, a_{L-1}$ be $L$ non-negative numbers. Create a list of
  size $2L$ as 
  \begin{equation*} 
    \bb \triangleq [0, a_0, 0, a_1, \dots, 0, a_{L-2}, a_{L-1}, +\infty ].
  \end{equation*}
  Feeding this list to the pruned radix-$2L$ sorter will result in an output
  list of the form 
  \begin{equation*}
    \textstyle
    [\underbrace{0, 0, \dots, 0}_{\text{$L-1$ zeros}}, a_{(0)}, a_{(1)}, \dots,
    a_{(L-1)}, +\infty]
  \end{equation*}
  where $a_{(0)} \le a_{(1)} \le \dots \le a_{(L-1)}$ is the ordered
  permutation of $a_0, a_1, \dots, a_{L-1}$.
\end{proposition}
\begin{IEEEproof}
  It is clear that the assumptions \eqref{eq:oldListSorted} and
  \eqref{eq:positiveIncrements} hold for $\bb$. The proof of
  Proposition~\ref{prop:prunedRadix} shows if the last element of the list is
  additionally known to be the largest element, the pruned radix-$2L$ sorter
  sorts the entire list. 
\end{IEEEproof}

Note that while the same comparator network of a pruned radix-$2L$ sorter is
used for sorting $L$ numbers, $L$ separate $L$-to-$1$ multiplexers are required
to output the sorted list.
\subsection{Latency of Metric Sorting} \label{sec:sortingLatency}
We assume that the sorting procedure is carried out in a single clock cycle. 
A decoder based on the full radix-$2L$ sorter, only needs to sort the path
metrics for the information indices, hence, the total sorting latency of such an
implementation is
\begin{equation}
  D_{\rm MS}(\calA) = \abs{\calA} =  NR~\text{cycles}.
  \label{eq:latencyR}
\end{equation}

Using the pruned radix-$2L$ sorter, additional sorting steps are required at the
end of each contiguous set of frozen indices. Let $F_C(\calA)$ denote the number
of \emph{clusters} of frozen bits for a given information set $\calA$.\footnote{
  More precisely we assume $\calF = \bigcup_{j=1}^{F_C(\calA)} \calF_j$ such
  that (i) $\calF_j \cap \calF_{j'} = \emptyset$ if $j \ne j'$, i.e.,
  $\{\calF_j: j=1,\dots,F_C(\calA)\}$ is a partition of $\calF$; (ii) for every
  $j$, $\calF_j$ is a contiguous subset of $\IndexSet{N}$; and (iii) for every
  pair $j \ne j'$, $\calF_j \cup \calF_{j'}$ is \emph{not} a contiguous subset
  of $\IndexSet{N}$. It can be easily checked that such a partition always
  exists and is unique.%
} The
metric sorting latency using the pruned radix-$2L$ sorter is then
\begin{equation}
  D_{\rm MS}(\calA) = \abs{\calA} + F_C(\calA) =  NR + F_C(\calA)~\text{cycles}.
  \label{eq:latencyPR}
\end{equation}

\section{Implementation Results}\label{sec:results}
In this section, we present synthesis results for our SCL decoder architecture. 
For fair comparison with~\cite{Lin15}, we use a TSMC $90~\text{nm}$ technology 
with a typical timing library ($1~\text{V}$ supply voltage, $25^{\circ}$C 
operating temperature) and our decoder of \cite{Bala14} is re-synthesized 
using this technology. All synthesis runs are performed with timing constraints
that are not achievable, in order to assess the maximum achievable operating
frequency of each design, as reported by the synthesis tool. For our synthesis
results, we have used $P = 64$ PEs per SC decoder core, as in
\cite{Leroux13,Bala14}. The hardware \emph{efficiency} is defined as the
throughput per unit area and it is measured in Mbps/mm$^2$. The decoding
throughput of all decoders is:
\begin{equation} 
  T_{\mathrm{SCL}}(N,P,\calA,f) = \frac{f \cdot
  N}{D_{\mathrm{SCL}}(N,P,\calA)},
\end{equation}
where $f$ is the operating frequency of the decoder.

We first compare the LLR-based decoder of this work with our previous LL-based
decoder \cite{Bala14}, in order to demonstrate the improvements obtained by
moving to an LLR-based formulation of SCL decoding. Then, we examine the effect
of using the pruned radix-$2L$ sorter on our LLR-based SCL decoder. Finally, we
compare our LLR-based decoder with the LL-based decoder of~\cite{Lin15} (since
\cite{Lin15} is an improved version of \cite{Lin14}, we do not compare directly
with \cite{Lin14}) and \cite{Yuan14}. A direct comparison with the SCL decoders
of \cite{Zhang14b,Lin14b} is unfortunately not possible, as the authors do not
report their synthesis results in terms of mm$^2$. Finally, we provide some
discussion on the effectiveness of a CA-SCLD. 

\subsection{Quantization Parameters}
\begin{figure}[tb]
  \centering
  \scalebox{0.95}{\begin{tikzpicture}[font=\footnotesize]
  \begin{semilogyaxis}[width=\columnwidth, height=0.375\textheight,%
      enlargelimits=false,
      xmin=1.5, xmax=4,
      xlabel={$E_b/N_0$ (dB)}, ylabel={FER}, grid=both,%
      xtick={0,0.5,1,1.5,2,2.5,3,3.5,4},
      legend pos=south west, 
      legend cell align=left
    ]
 
    \addplot+[green!50!black, dashed, mark=x, thick,
    error bars/.cd, x dir=none, y dir=both, y explicit relative]
    table[x index=0,y index = 1, y error index=2]{data/fixed/1024-0.5-1.dat};
    \label{plot:quant:scfixed}

    \addplot+[green!50!black, solid, mark=x, thick,
    error bars/.cd, x dir=none, y dir=both, y explicit relative]
    table[x index=0,y index = 1, y error index=2]{data/float/1024-0.5-1.dat};
    \label{plot:quant:scfloat}
    
    \addplot+[blue, dashed, mark=o, thick,
    error bars/.cd, x dir=none, y dir=both, y explicit relative]
    table[x index=0,y index = 1, y error index=2]{data/fixed/1024-0.5-2.dat};
    \label{plot:quant:l2fixed}
    
    \addplot+[blue, densely dotted, mark=o, thick,
    error bars/.cd, x dir=none, y dir=both, y explicit relative]
    table[x index=0,y index = 1, y error index=2]{data/fixed/1024-0.5-2-LL.dat};
    \label{plot:quant:l2fixedll}
 
    \addplot+[blue, solid, mark=o, thick,
    error bars/.cd, x dir=none, y dir=both, y explicit relative]
    table[x index=0,y index = 1, y error index=2]{data/float/1024-0.5-2.dat};
    \label{plot:quant:l2float}

    \addplot+[red, dashed, mark=square, thick,
    error bars/.cd, x dir=none, y dir=both, y explicit relative]
    table[x index=0,y index = 1, y error index=2]{data/fixed/1024-0.5-4.dat};
    \label{plot:quant:l4fixed}

    \addplot+[red, densely dotted, mark=square, thick,
    error bars/.cd, x dir=none, y dir=both, y explicit relative]
    table[x index=0,y index = 1, y error index=2]{data/fixed/1024-0.5-4-LL.dat};
    \label{plot:quant:l4fixedll}

    \addplot+[red, solid, mark=square, thick,
    error bars/.cd, x dir=none, y dir=both, y explicit relative]
    table[x index=0,y index = 1, y error index=2]{data/float/1024-0.5-4.dat};
    \label{plot:quant:l4float}

    \addplot+[teal, dashed, mark=triangle, thick,
    error bars/.cd, x dir=none, y dir=both, y explicit relative]
    table[x index=0,y index = 1, y error index=2]{data/fixed/1024-0.5-8.dat};
    \label{plot:quant:l8fixed}

    \addplot+[teal, densely dotted, mark=triangle, thick,
    error bars/.cd, x dir=none, y dir=both, y explicit relative]
    table[x index=0,y index = 1, y error index=2]{data/fixed/1024-0.5-8-LL.dat};
    \label{plot:quant:l8fixedll}
  
    \addplot+[teal, solid, mark=triangle, thick,
    error bars/.cd, x dir=none, y dir=both, y explicit relative]
    table[x index=0,y index = 1, y error index=2]{data/float/1024-0.5-8.dat};
    \label{plot:quant:l8float}
  \end{semilogyaxis}
  \node [draw,fill=white] at (rel axis cs: 0.0878911875,1.953125) {%
    \shortstack[l]{%
      \ref{plot:quant:scfixed} SC Decoder, $Q=6$\\
      \ref{plot:quant:scfloat} SC Decoder, Floating-Point \\
      \ref{plot:quant:l2fixed} $L=2$, LLR-based, $Q=6$\\
      \ref{plot:quant:l2fixedll} $L=2$, LL-based, $Q=4$ \cite{Bala14}\\
      \ref{plot:quant:l2float} $L=2$, Floating-Point
    }
  };
  \node [draw,fill=white] at (rel axis cs: -0.28152,1.25) {%
    \shortstack[l]{%
      \ref{plot:quant:l4fixed} $L=4$, LLR-based, $Q=6$\\
      \ref{plot:quant:l4fixedll} $L=4$, LL-based, $Q=4$ \cite{Bala14}\\
      \ref{plot:quant:l4float} $L=4$, Floating-Point\\
      \ref{plot:quant:l8fixed} $L=8$, LLR-based, $Q=6$\\
      \ref{plot:quant:l8fixedll} $L=8$, LL-based, $Q=4$ \cite{Bala14}\\
      \ref{plot:quant:l8float} $L=8$, Floating-Point
    }
  };
\end{tikzpicture}}
  \caption{The performance of floating-point vs. fixed-point SCL decoders.
    $M=8$ quantization bits are used for the path metric in fixed-point SCL
  decoders.}
  \label{fig:quant}
\end{figure}
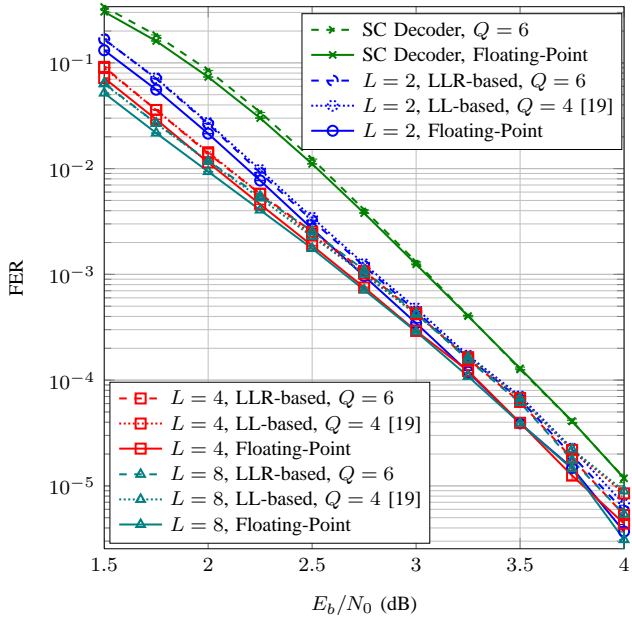

In Figure~\ref{fig:quant}, we present the FER of floating-point and fixed-point
implementations of an LL-based and an LLR-based SCL decoder for a $(1024,512)$
polar code as a function of SNR.\footnote{The code is
  optimized for ${E_b}/{N_0} = 2 \text{dB}$ and constructed using the
  Monte-Carlo method of \cite[Section IX]{Arik09}.
} For the floating-point simulations we have used the exact implementation of
the decoder, i.e., for computing the LLRs the update rule $f_-$ of
\eqref{eq:fminus} is used and the path metric is iteratively updated according
to \eqref{eq:pmupdate}. In contrast, for the fixed-point simulations we have
used the min-sum approximation of the decoder (i.e., replaced $f_-$ with
$\tilde{f}_-$ as in \eqref{eq:fminustilde}) and the approximated path metric
update rule of \eqref{eq:pmupdateApprox}.

We observe that the LL-based and the LLR-based SCL have practically
indistinguishable FER performance when quantizing the channel LLs and the
channel LLRs with $Q = 4$ bits and $Q = 6$ bits respectively. Moreover, in our
simulations we observe that the performance of the LL and the LLR-based SCL
decoder is degraded significantly when $Q < 6$ and $Q < 4$, respectively. As
discussed in Section~\ref{subsec:quantization}, metric quantization requires at
most $M = n + Q - 1$ bits. However, in practice, much fewer bits turn out to be
sufficient.  For example, in our simulations for $N = 1024$ and $Q = 6$, setting
$M = 8$ leads to the same performance as the worst-case $M = 15$, while setting
$M = 7$ results in a significant performance degradation due to metric
saturation.  Thus, all synthesis results of this section are obtained for $Q =
4$ for the LL-based decoder of \cite{Bala14}, and $Q = 6$ and $M = 8$ for the
LLR-based decoder for a fair (i.e., iso-FER) comparison. 

The authors of \cite{Yuan14} do not provide the FER curves for their fixed-point
implementation of SCLD and the authors of \cite{Lin15} only provide the FERs for
a CA-SCLD \cite[Figure~2]{Lin15}. Nevertheless, we assume their quantization
schemes will not result in a \emph{better} FER performance for a \emph{standard}
SCLD than that of \cite{Bala14} since they both implement exactly the same
algorithm as in \cite{Bala14} (using a different \emph{architecture} than
\cite{Bala14}),

\subsection{Gains due to LLR-based Formulation of SCL Decoding}
Our previous LL-based architecture of \cite{Bala14} and the LLR-based
architecture with a radix-$2L$ sorter presented in this paper are identical
except that the former uses LLs while the latter uses LLRs.  Therefore, by
comparing these two architectures we can specifically identify the improvements
in terms of area and decoding throughput that arise directly from the
reformulation of SCL decoding in the LLR domain.

The cycle count for our SCL decoder using the radix-$2L$ sorter when
decoding a $(1024,512)$ polar code is $D_{\mathrm{SCL}}(N,P,\calA) =
2592~\text{cycles}$ (see \eqref{eq:latency} and \eqref{eq:latencyR}).

\begin{table}[tb]
  \footnotesize
  \caption{Comparison with LL-based implementation}\label{tab:llvsllr}
  \centering 
  \begin{tabular}{l|ccc|ccc} 
    & \multicolumn{3}{c|}{LL-Based \cite{Bala14}}
    & \multicolumn{3}{c}{LLR-Based} \\
    \hline
    & $L = 2$ & $L = 4$ & $L = 8$ %
    & $L = 2$ & $L = 4$ & $L = 8$ %
    \\
    \hline
    Freq. (MHz)
    & $794$ & $730$ & $408$ %
    & $847$ & $758$ & $415$ %
    \\
    Lat. (Cyc./bit)
    & $2.53$ & $2.53$ & $2.53$ %
    & $2.53$ & $2.53$ & $2.53$ %
    \\
    T/P (Mbps)
    & $314$ & $288$ & $161$ %
    & $335$ & $299$ & $164$ %
    \\
    Area  (mm$^2$)
    & $1.38$ & $2.62$ & $5.38$ %
    & $0.88$ & $1.75$ & $3.87$ %
    \\
    \hline
    Efficiency 
    & $227$& $110$& $30$%
    & $380$& $171$& $42$%
  \end{tabular}
\end{table}

\begin{table}[tb] 
  \centering 	
  \caption{Cell Area Breakdown for the LL-Based and the Radix-2L LLR-based SCL
    Decoders ($R = \frac12$, $N=1024$)}
  \begin{tabular}{c|ccc} 
    & LL-Based \cite{Bala14} 	& LLR-Based		& Reduction \\
    \hline 
    List Size	& \multicolumn{3}{c}{$L=2$} \\ 
    \hline
    Total Area (mm$^2$)	
    & $1.38$			& $0.88$		& $36\%$ \\
    Memory (mm$^2$)					
    & $1.07$			& $0.80$		& $25\%$ \\
    MCU (mm$^2$)		
    & $0.28$			& $0.06$		& $79\%$\\%
    Metric Sorter (mm$^2$)	
    & $1.34 \times 10^{-3}$	& $0.75 \times 10^{-3}$ & $44\%$ \\
    Other (mm$^2$)					
    & $0.03$			& $0.02$ 		& $50\%$\\%
    \hline 
    List Size	& \multicolumn{3}{c}{$L=4$} \\ 
    \hline
    Total Area (mm$^2$)	
    & $2.62$			& $1.75$		& $33\%$\\%
    Memory (mm$^2$)					
    & $1.92$			& $1.57$		& $18\%$\\%
    MCU (mm$^2$)						
    & $0.54$			& $0.11$		& $80\%$\\%
    Metric Sorter (mm$^2$)	
    & $13.92 \times 10^{-3}$	& $9.23 \times 10^{-3}$	& $33\%$ \\
    Other (mm$^2$)					
    & $0.15$			& $0.06$		& $60\%$ \\
    \hline 
    List Size	& \multicolumn{3}{c}{$L=8$} \\ 
    \hline
    Total Area (mm$^2$)			
    & $5.38$			& $3.87$		& $28\%$\\ 
    Memory (mm$^2$)					
    & $4.08$			& $3.46$		& $15\%$\\ 
    MCU (mm$^2$)						
    & $0.82$			& $0.18$		& $78\%$\\ 
    Metric Sorter (mm$^2$) 	
    & $70.65 \times 10^{-3}$	& $54.05 \times 10^{-3}$& $24\%$\\ 
    Other (mm$^2$)	 				
    & $0.41$			& $0.18$ 		& $56\%$\\ 
    \hline 
  \end{tabular} 
  \label{tab:syntharea}
\end{table}
From Table~\ref{tab:llvsllr}, we see that our LLR-based SCL decoder occupies
$36\%$, $33\%$, and $28\%$ smaller area than our LL-based SCL decoder of
\cite{Bala14} for $L=2$, $L=4$, and $L=8$, respectively.  We present the area
breakdown of the LL-based and the LLR-based decoders in
Table~\ref{tab:syntharea} in order to identify where the area reduction mainly
comes from and why the relative reduction in area decreases with increasing list
size $L$.  The \emph{memory} area corresponds to the combined area of the LLR
(or LL) memory, the partial sum memory, and the path memory. We observe that, in
absolute terms, the most significant savings in terms of area come from the
memory, where the area is reduced by up to $0.62~\text{mm}^2$ for $L=8$. On the
other hand, in relative terms, the biggest savings in terms of area come from
the MCU with an average area reduction of $79\%$. The relative reduction in the
memory area decreases with increasing list size $L$.  This happens because each
bit-cell of the partial sum memory and the path memory contains $L$-to-$L$
crossbars, whose size grows quadratically with $L$, while the LL (and LLR)
memory grows only linearly in size with $L$. Thus, the the size of the partial
sum memory and the path memory, which are not affected by the LLR-based
reformulation, becomes more significant as the list size is increased, and the
relative reduction due to the LLR-based formulation is decreased. Similarly, the
relative reduction in the metric sorter area decreases with increasing $L$,
because the LLR-based formulation only decreases the bit-width of the $L(2L-1)$
comparators of the radix-$2L$ sorter but it does not affect the size of the
sorting logic, which dominates the sorter area as the list size is increased.

From Table~\ref{tab:llvsllr}, we observe that the operating frequency (and,
hence, the throughput) of our LLR-based decoder is $7\%$,
$3\%$, and $2\%$ higher than that of our LL-based SCL decoder of \cite{Bala14}
for $L=2$, $L=4$, and $L=8$, respectively.

Due to the aforementioned improvements in area and decoding throughput, the 
LLR-based reformulation of SCL decoding leads to hardware decoders with
$67\%$, $55\%$, and $40\%$ better hardware efficiency than the corresponding 
LL-based decoders of~\cite{Bala14}, for $L=2$, $L=4$, and $L = 8$, respectively.

\subsection{Radix-$2L$ Sorter versus Pruned Radix-$2L$ Sorter}
\label{subsec:r2lvspr2l}
One may expect the pruned radix-$2L$ sorter to always outperform the radix-$2L$
sorter. However, the decoder equipped with the pruned radix-$2L$ sorter needs to
stall slightly more often to perform the additional sorting steps after groups
of frozen bits. In particular, a $(1024,512)$ polar code contains
$F_C(\calA) = 57$ groups of frozen bits. Therefore, the total sorting latency
for the pruned radix-$2L$ sorter is $D_{\rm MS}(\calA) = \abs{\calA} +
F_C(\calA) = 569$ cycles (see \eqref{eq:latencyPR}). Thus, we have
$D_{\mathrm{SCL}}(N,P,\calA) = 2649$ cycles, which is an increase of
approximately $2\%$ compared to the decoder equipped with a full radix-$2L$
sorter. Therefore, if using the pruned radix-$2L$ does not lead to a more than
$2\%$ higher clock frequency, the decoding throughput will actually be reduced.
\begin{table}[tb]
  \footnotesize
  \caption{Radix-$2L$ vs. Pruned Radix-$2L$ Sorter}
  \label{tab:radix2lvsprunedradix2l} 
  \begin{tabular}{l|ccc|ccc} 
    & \multicolumn{3}{c|}{Radix-$2L$ Sorter}
    & \multicolumn{3}{c}{Pruned Radix-$2L$ Sorter} \\ 
    \hline
    & $L = 2$ & $L = 4$ & $L = 8$ %
    & $L = 2$ & $L = 4$ & $L = 8$ %
    \\
    \hline
    Freq. (MHz)
    & $847$ & $758$ & $415$ %
    & $848$ & $794$ & $637$ %
    \\
    Lat. (Cyc./bit)
    & $2.53$ & $2.53$ & $2.53$ %
    & $2.59$ & $2.59$ & $2.59$ %
    \\
    T/P (Mbps)
    & $335$ & $299$ & $164$ %
    & $328$ & $307$ & $246$ %
    \\
    Area  (mm$^2$)
    & $0.88$ & $1.75$ & $3.87$ %
    & $0.9$ & $1.78$ & $3.85$ %
    \\
    \hline
    Efficiency
    & $380$& $171$& $42$%
    & $364$& $172$& $64$%
  \end{tabular}
\end{table}

As can be observed in Table~\ref{tab:radix2lvsprunedradix2l}, this is exactly
the case for $L=2$, where the LLR-based SCL decoder with the pruned radix-2L
sorter has a $2\%$ \emph{lower} throughput than the LLR-based SCL decoder with
the full radix-$2L$ sorter. However, for $L \geq 4$ the metric sorter starts to
lie on the critical path of the decoder and therefore using the pruned
radix-$2L$ sorter results in a significant increase in throughput of up to
$50\%$ for $L = 8$.

To provide more insight into the effect of the metric sorter on our SCL
decoder, in Table~\ref{tab:msdelay} we present the metric sorter delay and the
critical path start- and endpoints of each decoder of
Table~\ref{tab:radix2lvsprunedradix2l}. The critical paths for $L = 2$ 
and $L=4,8,$ are also annotated in Figure~\ref{fig:scldecarch} with green dashed 
lines and red dotted lines, respectively. We denote the register of the controller
which stores the internal LLR memory read address by $R_{\text{IM}}$. Moreover,
let $D_{\hat{U}_s}$ and $D_{M}$ denote a register of the partial sum memory and
the metric memory, respectively.  From Table~\ref{tab:msdelay}, we observe that,
for $L=2$, the radix-$2L$ sorter does not lie on the critical path of the
decoder, which explains why using the pruned radix-$2L$ sorter does not improve
the operating frequency of the decoder. 
For $L \geq 4$ the metric sorter does lie on the critical path of the decoder
and using the pruned radix-$2L$ sorter results in a significant increase in
the operating frequency of up to $53\%$. It is interesting to note
that using the pruned radix-$2L$ sorter eliminates the metric sorter
completely from the critical path of the decoder for $L=4$. For $L=8$, even the
pruned radix-$2L$ sorter lies on the critical path of the decoder, but the
delay through the sorter is reduced by $40\%$.

\begin{table}[tb] 
  \begin{threeparttable} 
    \caption{Metric Sorter Delay and Critical Path Start- and Endpoints for our
      LLR-Based SCL Decoder Using the Radix-$2L$ and the Pruned Radix-$2L$
    Sorters.} 
    \label{tab:msdelay} 
    \begin{tabular}{l|ccc|ccc}  
      & \multicolumn{3}{c|}{Radix-$2L$ Sorter} 
      & \multicolumn{3}{c}{Pruned Radix-$2L$ Sorter} \\ 
      & $L = 2$	& $L = 4$ & $L = 8$ 
      & $L = 2$	& $L = 4$ & $L = 8$\\ 
      \hline 
      Delay (ns)		
      & $0.50$\tnote{a}	& $0.80$ & $1.83$ 
      & $0.50$\tnote{a}	& $0.54$ & $1.09$ \\ \hline 
      CP Startpoint	
      & $R_{\text{IM}}$	& $D_{M}$&	$D_{M}$	
      & $R_{\text{IM}}$	& $R_{\text{IM}}$	&	$D_{M}$\\ 
      CP Endpoint		
      & $D_{M}$		& $D_{\hat{U}_s}$	& $D_{\hat{U}_s}$ 
      & $D_{M}$		& $D_{M}$		& $D_{\hat{U}_s}$ 
    \end{tabular} 
    \begin{tablenotes} 
    \item [a] Note that the true delay of the pruned radix-$2L$ sorter is always
      smaller than the delay of the radix-$2L$ sorter. However, for $L=2$, both
      sorters meet the synthesis timing constraint, which was set to $0.50$ ns.
    \end{tablenotes}
  \end{threeparttable}
\end{table}

\subsection{Comparison with LL-based SCL Decoders} \label{sec:compareOther} 
In Table~\ref{tab:scldecoders}, we compare our LLR-based decoder with the
LL-based decoders of~\cite{Lin15} and~\cite{Yuan14} along with our LL-based
decoder of~\cite{Bala14}. For the comparisons, we pick our SCL decoder with the
best hardware efficiency for each list size, i.e., for $L=2$ we pick the SCL
decoder with the radix-$2L$ sorter, while for $L=4,8,$ we pick the SCL decoder
with the pruned radix-$2L$ sorter. Moreover, we pick the decoders with the
best hardware efficiency from~\cite{Yuan14}, i.e., the $4b$-rSCL decoders.

\begin{table*}[!t]
  \begin{threeparttable}
    \caption{SCL Decoder  Synthesis Results ($R = \frac12$, $N=1024$)}
    \label{tab:scldecoders} 
    \centering 
    \begin{tabular}{l|ccc|ccc|ccc|cc|cc}
      & \multicolumn{3}{c|}{LLR-Based} 
      & \multicolumn{3}{c|}{LL-Based \cite{Bala14}}
      & \multicolumn{3}{c|}{LL-Based \cite{Lin15}\tnote{a}}
      & \multicolumn{4}{c}{LL-Based \cite{Yuan14}\tnote{b}}\\
      \hline 
      & $L = 2$ & $L = 4$ & $L = 8$%
      & $L = 2$ & $L = 4$ & $L = 8$%
      & $L = 2$ & $L = 4$ & $L = 8$ 
      & $L = 2$ & $L = 4$ & $L = 2$ & $L = 4$\\
      \hline 
      Technology
      & \multicolumn{3}{c|}{TSMC 90nm}
      & \multicolumn{3}{c|}{TSMC 90nm}
      & \multicolumn{3}{c|}{TSMC 90nm}
      & \multicolumn{2}{c|}{Scaled to 90nm\tnote{c}}
      & \multicolumn{2}{c}{ST 65nm} \\
      \hline
      Freq. (MHz)
      & $847$ & $794$ & $637$ %
      & $794$ & $730$ & $408$ %
      & $507$ & $492$ & $462$ %
      & $361$ & $289$ & $500$  & $400$ \\
      Lat. (Cycles/bit)
      & $2.53$ & $2.59$ & $2.59$ %
      & $2.53$ & $2.53$ & $2.53$ %
      & $2.53$ & $2.53$ & $3.03$ %
      & $1.00$ & $1.00$ & $1.00$  & $1.00$ \\
      T/P (Mbps)
      & $335$ & $307$ & $246$ %
      & $314$ & $288$ & $161$ %
      & $200$ & $194$ & $153$ %
      & $362$ & $290$ & $501$ & $401$ \\
      Area  (mm$^2$)
      & $0.88$ & $1.78$ & $3.58$ %
      & $1.38$ & $2.62$ & $5.38$ %
      & $1.23$ & $2.46$ & $5.28$ %
      & $2.03$ & $4.10$ & $1.06$ & $2.14$ \\ 
      \hline
      Efficiency 
      & $380$& $172$& $69$%
      & $227$& $110$& $30$%
      & $163$& $79$& $29$%
      & $178$& $71$& $473$ & $187$
    \end{tabular}
    \begin{tablenotes}
    \item [a] The synthesis results in \cite{Lin15} are provided with up to
      $16$ PEs per path. The reported numbers in this table are the
      corresponding synthesis results using $64$ PEs per path and are courtesy
      of the authors of \cite{Lin15}.
    \item [b] The authors of \cite{Yuan14} use $3$ quantization bits for the
      channel LLs and a tree SC architecture, while \cite{Bala14,Lin15} use $4$
      quantization bits for the channel LLs and a semi-parallel architecture
      with $P = 64$ PEs per path.
    \item [c] We use the standard assumption that area scales as $s^2$ and
      frequency scales as $1/s$, where $s$ is the feature size.
    \end{tablenotes}
  \end{threeparttable}
\end{table*}
\subsubsection{Comparison with \cite{Lin15}} 
From Table~\ref{tab:scldecoders} we observe that our LLR-based SCL
decoder has an approximately $28\%$ smaller area than the LL-based SCL
decoder of \cite{Lin15} for all list sizes. 
Moreover, the throughput of our LLR-based SCL decoder is up to $70\%$ 
higher than the throughput achieved by the LL-based SCL decoder of~\cite{Lin15},
leading to a $137\%$, $118\%$, and $120\%$ better hardware efficiency for $L=2$,
$L=4$ and $L=8$, respectively.

\subsubsection{Comparison with \cite{Yuan14}}
The synthesis results of~\cite{Yuan14} are given for a $65$nm technology, which
makes a fair comparison difficult. Nevertheless, in order to enable as fair a
comparison as possible, we scale the area and the frequency to a $90$nm
technology in Table~\ref{tab:scldecoders} (we have also included the original
results for completeness). Moreover, the authors of~\cite{Yuan14} only provide
synthesis results for $L=2$ and $L=4$.  In terms of area, we observe that our
decoder is approximately $57\%$ smaller than the decoder of~\cite{Yuan14} for
all list sizes. We also observe that for $L=2$ our decoder has a $7\%$ lower
throughput than the decoder of~\cite{Yuan14}, but for $L=4$ the throughput of
our decoder is $6\%$ higher than that of~\cite{Yuan14}. Overall, the hardware
efficiency of our LLR-based SCL decoder is $115\%$ and $142\%$ better than that
of \cite{Yuan14} for $L=2$ and $L=4$ respectively. 

\subsection{CRC-Aided SCL Decoder} \label{sec:CASCLDresults}
As discussed in Section~\ref{sec:CRCAidedSCLD}, the performance of the SCL
decoder can be significantly improved if it is assisted for its final choice by
means of a CRC which rejects some incorrect codewords from the final set of $L$
candidates. However, there is a trade-off between the length of the CRC and the
performance gain. A longer CRC, rejects more incorrect codewords but, at
the same time, it degrades the performance of the inner polar code by increasing
its rate. Hence, the CRC improves the overall performance if the performance
degradation of the inner polar code is compensated by rejecting the incorrect
codewords in the final list.

\subsubsection{Choice of CRC}
We picked three different CRCs  of lengths $r=4$, $r=8$ and
$r=16$ from \cite{wiki:CRC} with generator polynomials:
\begin{subequations}
  \begin{align}
    g(x) & = x^4 + x + 1, \label{eq:CRC4gen} \\
    g(x) & = x^8 + x^7 + x^6 + x^4 + x^2 + 1, \text{ and} \label{eq:CRC8gen} \\
    g(x) & = x^{16} + x^{15} + x^{2} + 1, \label{eq:CRC16gen}
  \end{align}
\end{subequations}
respectively and evaluated the empirical performance of the SCL decoders of list
sizes of $L=2$, $L=4$, $L=8$, aided by each of these three CRCs in
the regime of $E_b/N_0 = 1.5~\text{dB}$ to $E_b/N_0=4~\text{dB}$. 

For $L=2$ it turns out that the smallest CRC, represented by the
generator polynomial in \eqref{eq:CRC4gen}, is the best choice. Using longer
CRCs at $E_b/N_0 \le 3~\text{dB}$, the performance degradation of the polar code
is dominant, causing the CRC-aided SCL decoder to perform \emph{worse}
than the standard SCL decoder. Furthermore, at higher SNRs, longer CRCs do
not lead to a significantly better performance than the CRC-$4$.

For $L=4$, allocating $r=8$ bits for the CRC of \eqref{eq:CRC8gen} turns out to
be the most beneficial option. CRC-$4$ and CRC-$8$ will lead to almost identical
FER at $E_b/N_0 < 2.25$ dB while CRC-$8$ improves the FER significantly more
than CRC-$4$ at higher SNRs. Furthermore, CRC-$16$ leads to the same performance
as CRC-$8$ at high SNRs and worse performance than CRC-$8$ in low-SNR regime.

Finally, for $L=8$ we observe that CRC-$16$ of \eqref{eq:CRC16gen} is the best
candidate among the three different CRCs in the sense that the performance of
the CRC-aided SCL decoder which uses this CRC is significantly better than that
of the decoders using CRC-$4$ or CRC-$8$ for $E_b/N_0 > 2.5~\text{dB}$, while
all three decoders have almost the same FER at lower SNRs (and they all perform
better than a standard SCL decoder). 

\begin{figure}[htb]
  \centering
  \subfloat[$L=2$]{\scalebox{0.95}{\begin{tikzpicture}[font=\footnotesize]
  \begin{semilogyaxis}[width=\columnwidth, height=0.25\textheight,%
      enlargelimits=false,
      xmin=1.5, xmax=4, 
      xlabel={$E_b/N_0$ (dB)}, ylabel={FER}, grid=both,%
    legend pos=north east, legend cell align=left]
    
    \addplot+[blue, dashed, mark=o, thick,
    error bars/.cd, x dir=none, y dir=both, y explicit relative]
    table[x index=0,y index = 1, y error index=2]{data/fixed/1024-0.5-2.dat};
    \label{plot:crcl2:fixed}

    \addplot+[blue, solid, mark=o, thick,
    error bars/.cd, x dir=none, y dir=both, y explicit relative]
    table[x index=0,y index = 1, y error index=2]{data/float/1024-0.5-2.dat};
    \label{plot:crcl2:float}

    \addplot+[black, solid, mark=oplus, thick,
    error bars/.cd, x dir=none, y dir=both, y explicit relative]
    table[x index=0, y index = 1, y error index=2]
    {data/float/1024-0.5-2-CRC8.dat};
    \label{plot:crcl2:crc8}

    \addplot+[black, solid, mark=otimes*, mark options={fill=gray}, thick,
    error bars/.cd, x dir=none, y dir=both, y explicit relative]
    table[x index=0, y index = 1, y error index=2]
    {data/float/1024-0.5-2-CRC16.dat};
    \label{plot:crcl2:crc16}
 
    \addplot+[blue, dashed, mark=*, mark options={blue}, thick,
    error bars/.cd, x dir=none, y dir=both, y explicit relative]
    table[x index=0,y index = 1, y error index=2]
    {data/fixed/1024-0.5-2-CRC4.dat};
    \label{plot:crcl2:crc4fixed}
    
    \addplot+[blue, solid, mark=*, mark options={blue}, thick,
    error bars/.cd, x dir=none, y dir=both, y explicit relative]
    table[x index=0,y index = 1, y error index=2]
    {data/float/1024-0.5-2-CRC4.dat};
    \label{plot:crcl2:crc4float}

   \end{semilogyaxis}
   \node[draw, fill=white] at (rel axis cs: 0.0625,2.0) {%
     \shortstack[l]{%
       \ref{plot:crcl2:crc16} SCLD + CRC-$16$, Floating-Point \\
       \ref{plot:crcl2:crc8} SCLD + CRC-$8$, Floating-Point
     }
   };
   \node[draw, fill=white] at (rel axis cs: -0.265625,1.3125) {%
     \shortstack[l]{%
       \ref{plot:crcl2:fixed} SCLD, $Q=6$ \\
       \ref{plot:crcl2:float} SCLD, Floating-Point \\
       \ref{plot:crcl2:crc4fixed} SCLD + CRC-$4$, $Q=6$ \\
       \ref{plot:crcl2:crc4float} SCLD + CRC-$4$, Floating-Point
     }
   };
\end{tikzpicture}} \label{fig:crc2}}
  \\
  \subfloat[$L=4$]{\scalebox{0.95}{\begin{tikzpicture}[font=\footnotesize]
  \begin{semilogyaxis}[width=\columnwidth, height=0.26875\textheight,%
      enlargelimits=false,
      xmin=1.5, xmax=4, 
      xlabel={$E_b/N_0$ (dB)}, ylabel={FER}, grid=both,%
    legend pos=north east, legend cell align=left]
    
    \addplot+[red, dashed, mark=square, thick,
    error bars/.cd, x dir=none, y dir=both, y explicit relative]
    table[x index=0,y index = 1, y error index=2]{data/fixed/1024-0.5-4.dat};
    \label{plot:crcl4:fixed}

    \addplot+[red, solid, mark=square, thick,
    error bars/.cd, x dir=none, y dir=both, y explicit relative]
    table[x index=0,y index = 1, y error index=2]{data/float/1024-0.5-4.dat};
    \label{plot:crcl4:float}

    \addplot+[black, solid, mark=diamond, thick,
    error bars/.cd, x dir=none, y dir=both, y explicit relative]
    table[x index=0, y index = 1, y error index=2]
    {data/float/1024-0.5-4-CRC4.dat};
    \label{plot:crcl4:crc4}

    \addplot+[black, solid, mark=diamond*, mark options={fill=black}, thick,
    error bars/.cd, x dir=none, y dir=both, y explicit relative]
    table[x index=0, y index = 1, y error index=2]
    {data/float/1024-0.5-4-CRC16.dat};
    \label{plot:crcl4:crc16}
 
    \addplot+[red, dashed, mark=square*, mark options={red}, thick,
    error bars/.cd, x dir=none, y dir=both, y explicit relative]
    table[x index=0,y index = 1, y error index=2]
    {data/fixed/1024-0.5-4-CRC8.dat};
    \label{plot:crcl4:crc8fixed}
    
    \addplot+[red, solid, mark=square*, mark options={red}, thick,
    error bars/.cd, x dir=none, y dir=both, y explicit relative]
    table[x index=0,y index = 1, y error index=2]
    {data/float/1024-0.5-4-CRC8.dat};
    \label{plot:crcl4:crc8float}

   \end{semilogyaxis}
   \node[draw, fill=white] at (rel axis cs: 0.0625,2.046875) {%
     \shortstack[l]{%
       \ref{plot:crcl4:crc4} SCLD + CRC-$4$, Floating-Point \\
       \ref{plot:crcl4:crc16} SCLD + CRC-$16$, Floating-Point
     }
   };
   \node[draw, fill=white] at (rel axis cs: -0.265625,1.3125) {%
     \shortstack[l]{%
       \ref{plot:crcl4:fixed} SCLD, $Q=6$ \\
       \ref{plot:crcl4:float} SCLD, Floating-Point \\
       \ref{plot:crcl4:crc8fixed} SCLD + CRC-$8$, $Q=6$ \\
       \ref{plot:crcl4:crc8float} SCLD + CRC-$8$, Floating-Point
     }
   };
\end{tikzpicture}} \label{fig:crc4}}
  \\
  \subfloat[$L=8$]{\scalebox{0.95}{\begin{tikzpicture}[font=\footnotesize]
  \begin{semilogyaxis}[width=\columnwidth, height=0.275\textheight,%
      enlargelimits=false,
      xmin=1.5, xmax=4, 
      xlabel={$E_b/N_0$ (dB)}, ylabel={FER}, grid=both,%
    legend pos=north east, legend cell align=left]
    
    \addplot+[teal, dashed, mark=triangle, thick,
    error bars/.cd, x dir=none, y dir=both, y explicit relative]
    table[x index=0,y index = 1, y error index=2]{data/fixed/1024-0.5-8.dat};
    \label{plot:crcl8:fixed}

    \addplot+[teal, solid, mark=triangle, thick,
    error bars/.cd, x dir=none, y dir=both, y explicit relative]
    table[x index=0,y index = 1, y error index=2]{data/float/1024-0.5-8.dat};
    \label{plot:crcl8:float}

    \addplot+[black, solid, mark=pentagon, thick,
    error bars/.cd, x dir=none, y dir=both, y explicit relative]
    table[x index=0, y index = 1, y error index=2]
    {data/float/1024-0.5-8-CRC4.dat};
    \label{plot:crcl8:crc4}

    \addplot+[black, solid, mark=pentagon*, mark options={fill=black}, thick,
    error bars/.cd, x dir=none, y dir=both, y explicit relative]
    table[x index=0, y index = 1, y error index=2]
    {data/float/1024-0.5-8-CRC8.dat};
    \label{plot:crcl8:crc8}
 
    \addplot+[teal, dashed, mark=triangle*, mark options={teal}, thick,
    error bars/.cd, x dir=none, y dir=both, y explicit relative]
    table[x index=0,y index = 1, y error index=2]
    {data/fixed/1024-0.5-8-CRC16.dat};
    \label{plot:crcl8:crc16fixed}
    
    \addplot+[teal, solid, mark=triangle*, mark options={teal}, thick,
    error bars/.cd, x dir=none, y dir=both, y explicit relative]
    table[x index=0,y index = 1, y error index=2]
    {data/float/1024-0.5-8-CRC16.dat};
    \label{plot:crcl8:crc16float}

   \end{semilogyaxis}
   \node[draw, fill=white] at (rel axis cs: 0.0703125,2.0390625) {%
     \shortstack[l]{%
       \ref{plot:crcl8:crc4} SCLD + CRC-$4$, Floating-Point\\
       \ref{plot:crcl8:crc8} SCLD + CRC-$8$, Floating-Point
     }
   };
   \node[draw, fill=white] at (rel axis cs: -0.25,1.296875) {%
     \shortstack[l]{%
       \ref{plot:crcl8:fixed} SCLD, $Q=6$ \\
       \ref{plot:crcl8:float} SCLD, Floating-Point \\
       \ref{plot:crcl8:crc16fixed} SCLD + CRC-$16$, $Q=6$ \\
       \ref{plot:crcl8:crc16float} SCLD + CRC-$16$, Floating-Point
     }
   };
\end{tikzpicture}} \label{fig:crc8}}
  \caption{The performance of LLR-based SCL decoders compared to that of
    CRC-Aided SCL decoders for $L=2,4,8$. $M=8$ quantization bits are used for
    the path metric in fixed-point simulations.}
  \label{fig:crc}
\end{figure}
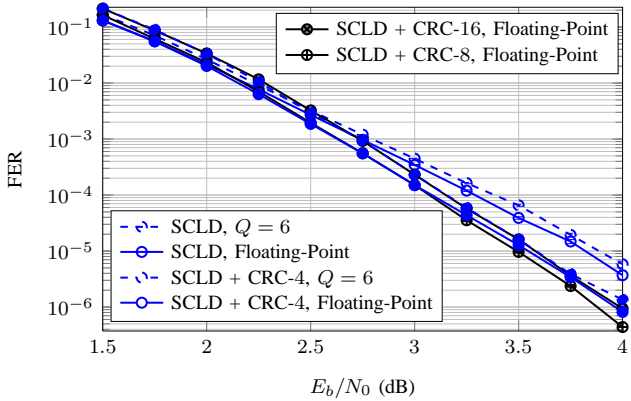
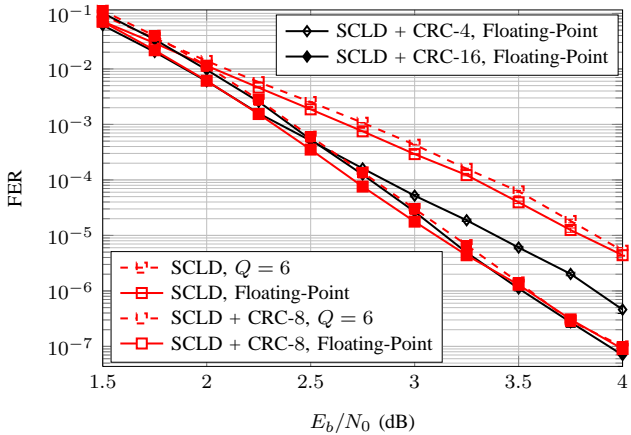
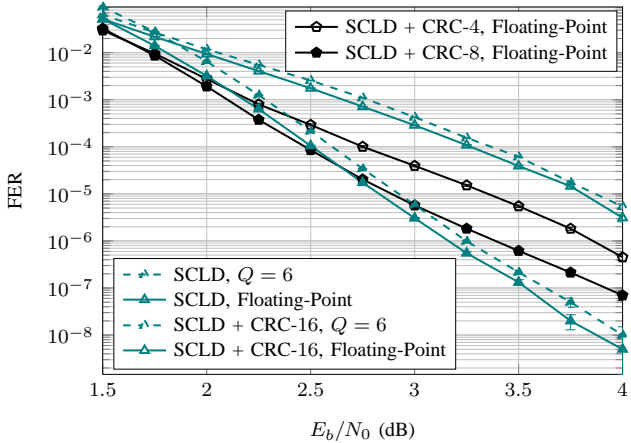

In Figure~\ref{fig:crc}, we compare the FER of the SCL decoder with that of the
CA-SCLD for list sizes of $L=2$, $L=4$ and $L=8$, using the above-mentioned
CRCs. We observe that the CRC-aided SCL decoders perform significantly better
than the standard SCL decoders. 

\subsubsection{Throughput Reduction}
Adding $r$ bits of CRC increases the number of information bits by $r$, while
reducing the number of groups of frozen channels by \emph{at most} $r$. As a
result, the sorting latency is generally increased, resulting in a decrease in
the throughput of the decoder. In Table~\ref{tab:throughput} we have computed
this decrease in the throughput for different decoders and we see that the
CRC-aided SCL decoders have slightly (at most $0.4\%$) reduced throughput. For
this table, we have picked the best decoder at each list size in terms of 
hardware efficiency from Table~\ref{tab:radix2lvsprunedradix2l}. 

\begin{table}[tb]
  \centering
  \caption{Throughput Reduction in CRC-Aided SCL Decoders}
  \label{tab:throughput}
  \begin{tabular}{c|r|ccc} 
    \multicolumn{2}{c|}{}
    & $L=2$ & $L=4$ & $L=8$ \\
    \hline
    \multicolumn{2}{r|}{Freq. (MHz)} & 
    $847$  & $794$ & $637$ \\
    \hline
    \multirow{4}{*}{SCLD} 
    & $\abs{\calA}$ 	& $512$ & $512$ & $512$\\
    & $F_C(\calA)$ 		& $57$ 	& $57$ 	& $57$ \\
    & Lat. (Cycles) 	& $2592$& $2649$& $2649$ \\ 
    & T/P (Mbits/s) 	& $335$ & $307$ & $246$ \\
    \hline
    \multirow{4}{*}{CA-SCLD} 
    & $\abs{\calA}$ 	& $516$ & $520$ & $528$\\
    & $F_C(\calA)$ 		& $55$  & $54$  & $52$ \\
    & Lat. (Cycles)     & $2596$& $2654$& $2660$\\
    & T/P (Mbits/s)     & $334$ & $306$ & $245$\\
    \hline
    \multicolumn{2}{r|}{Reduction (\%)} & 
    $0.2$ & $0.2$ & $0.4$ \\
    \hline
  \end{tabular}
\end{table}

\subsubsection{Effectiveness of CRC}
The area of the CRC unit for all synthesized decoders is in less than
$1~\mu\mathrm{m}^2$ for the employed TSMC $90$~nm technology. Moreover, the
CRC unit does not lie on the critical path of the decoder. Therefore, it does
not affect the maximum achievable operating frequency. Thus the incorporation of
a CRC unit is a highly effective method of improving the performance of an SCL
decoder. For example, it is interesting to note that the CA-SCLD with
$L=2$ has a somewhat lower FER than the standard SCL decoder with $L=8$ (in
both floating-point and fixed-point versions) in the regime of $E_b/N_0 >
2.5~\text{dB}$. Therefore, if a FER in the range of $10^{-3}$ to $10^{-6}$ is
required by the application, using a CA-SCLD with list size $L=2$ is preferable
to a standard SCL decoder with list size $L=8$ as the former has more than five
times higher hardware efficiency. 

\section{Discussion}\label{sec:discussion}
\subsection{SC Decoding or SCL Decoding?}
Modern communication standards sometimes allow very long block-lengths to be
used. The error-rate performance of polar codes under conventional SC decoding
is significantly improved if the block-length is increased. However, a long
block-length implies long decoding latency and large decoders. Thus, an
interesting question is whether it is better to use a long polar code with SC
decoding or a shorter one with SCL decoding, for a given target block-error
probability.  In order to answer this question, we first need to find some pairs
of short and long polar codes which have approximately the same block-error
probability under SCL and SC decoding, respectively to carry out a fair
comparison.

\begin{figure}[t!]
  \centering
  \subfloat[$(2048,1024)$ polar code under SC decoding versus $(1024,512)$
  modified polar code under CA-SCLD with $L=2$ and CRC-$4$ with generator
  polynomial \eqref{eq:CRC4gen}]{\scalebox{0.95}{\begin{tikzpicture}[font=\footnotesize]
  \begin{semilogyaxis}[width=\columnwidth, height=0.25\textheight,%
      enlargelimits=false,
      xmin=1.5, xmax=4, 
      xlabel={$E_b/N_0$ (dB)}, ylabel={FER}, grid=both,%
    legend pos=north east, legend cell align=left]
    
    \addplot+[blue, dashed, mark=*, mark options={blue}, thick,
    error bars/.cd, x dir=none, y dir=both, y explicit relative]
    table[x index=0,y index = 1, y error index=2]
    {data/fixed/1024-0.5-2-CRC4.dat};
    \label{plot:scvsscl2k:crc4fixed}

    \addplot+[blue, solid, mark=*, mark options={blue}, thick,
    error bars/.cd, x dir=none, y dir=both, y explicit relative]
    table[x index=0,y index = 1, y error index=2]
    {data/float/1024-0.5-2-CRC4.dat};
    \label{plot:scvsscl2k:crc4float}


    \addplot+[green!50!black, dashed, mark=star, thick,
    error bars/.cd, x dir=none, y dir=both, y explicit relative]
    table[x index=0,y index = 1, y error index=2]
    {data/fixed/2048-0.5-1.dat};
    \label{plot:scvsscl2k:scfixed}

    \addplot+[green!50!black, solid, mark=star,  thick,
    error bars/.cd, x dir=none, y dir=both, y explicit relative]
    table[x index=0,y index = 1, y error index=2]
    {data/float/2048-0.5-1.dat};
    \label{plot:scvsscl2k:scfloat}

   \end{semilogyaxis}
   \node[draw, fill=white] at (rel axis cs: 0.078125,1.953125) {%
     \shortstack[l]{%
       \ref{plot:scvsscl2k:scfixed} $N=2048$, SC, $Q=6$ \\
       \ref{plot:scvsscl2k:scfloat} $N=2048$, SC, Floating Point
     }
   };
   \node[draw, fill=white] at (rel axis cs: -0.203125,1.171875) {%
     \shortstack[l]{%
       \ref{plot:scvsscl2k:crc4fixed} $N=1024$, CA-SCLD, $Q=6$, $M=8$ \\
       \ref{plot:scvsscl2k:crc4float} $N=1024$, CA-SCLD, Floating-Point
     }
   };
\end{tikzpicture}}
  \label{fig:scvsscl2k}}
  \\
  \subfloat[$(4096,2048)$ polar code under SC decoding versus $(1024,512)$
  modified polar code under CA-SCLD with $L=4$ and CRC-$8$ with generator
  polynomial \eqref{eq:CRC8gen}]{\scalebox{0.95}{\begin{tikzpicture}[font=\footnotesize]
  \begin{semilogyaxis}[width=\columnwidth, height=0.26875\textheight,%
      enlargelimits=false,
      xmin=1.5, xmax=4, 
      xlabel={$E_b/N_0$ (dB)}, ylabel={FER}, grid=both,%
    legend pos=north east, legend cell align=left]
    
    \addplot+[red, dashed, mark=square*, mark options={red}, thick,
    error bars/.cd, x dir=none, y dir=both, y explicit relative]
    table[x index=0,y index = 1, y error index=2]
    {data/fixed/1024-0.5-4-CRC8.dat};
    \label{plot:scvsscl4k:crc8fixed}

    \addplot+[red, solid, mark=square*, mark options={red}, thick,
    error bars/.cd, x dir=none, y dir=both, y explicit relative]
    table[x index=0,y index = 1, y error index=2]
    {data/float/1024-0.5-4-CRC8.dat};
    \label{plot:scvsscl4k:crc8float}


    \addplot+[green!50!black, dashed, mark=asterisk, thick,
    error bars/.cd, x dir=none, y dir=both, y explicit relative]
    table[x index=0,y index = 1, y error index=2]
    {data/fixed/4096-0.5-1.dat};
    \label{plot:scvsscl4k:scfixed}

    \addplot+[green!50!black, solid, mark=asterisk,  thick,
    error bars/.cd, x dir=none, y dir=both, y explicit relative]
    table[x index=0,y index = 1, y error index=2]
    {data/float/4096-0.5-1.dat};
    \label{plot:scvsscl4k:scfloat}

   \end{semilogyaxis}
   \node[draw, fill=white] at (rel axis cs: 0.078125,1.96875) {%
     \shortstack[l]{%
       \ref{plot:scvsscl4k:scfixed} $N=4096$, SC, $Q=6$ \\
       \ref{plot:scvsscl4k:scfloat} $N=4096$, SC, Floating Point
     }
   };
   \node[draw, fill=white] at (rel axis cs: -0.203125,1.15625) {%
     \shortstack[l]{%
       \ref{plot:scvsscl4k:crc8fixed} $N=1024$, CA-SCLD, $Q=6$, $M=8$ \\
       \ref{plot:scvsscl4k:crc8float} $N=1024$, CA-SCLD, Floating-Point
     }
   };
\end{tikzpicture}}
  \label{fig:scvsscl4k}}
  \caption{CA-SCLD with $L = 2,4$, results in the same performance at
    block-length $N=1024$ as the conventional SC decoding with $N=2048$ and
  $N=4096$, respectively.}
  \label{fig:scvsscl}
\end{figure}
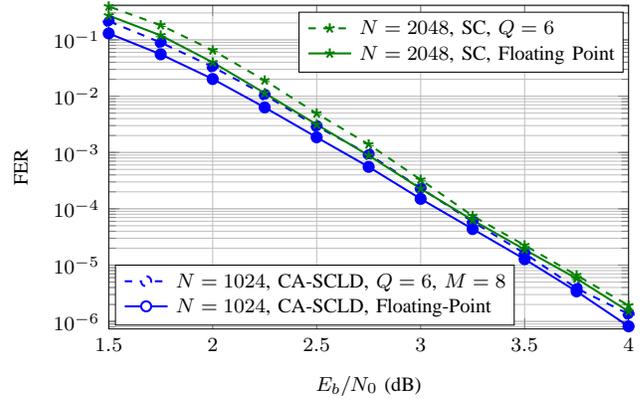
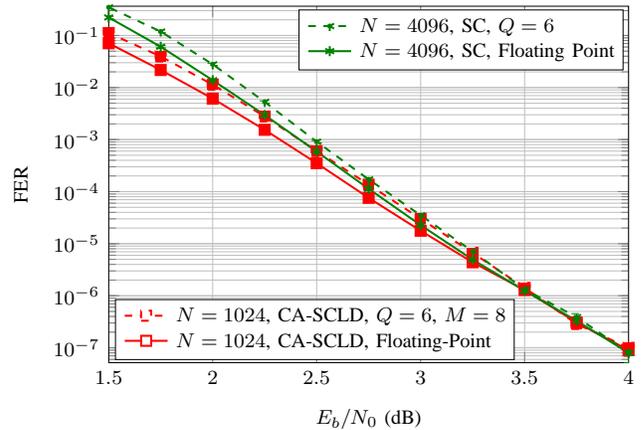

In Figure~\ref{fig:scvsscl2k} we see that a $(2048,1024)$ polar
code has almost the same block-error probability under SC decoding as a
$(1024,512)$ modified polar code under CA-SCLD with list
size $L=2$ and CRC-$4$ of \eqref{eq:CRC4gen}.  Similarly, in
Figure~\ref{fig:scvsscl4k} we see that a $(4096,2048)$ polar code has almost the
same block-error probability under SC decoding as an $(1024,512)$
modified polar code decoded under CA-SCLD with list size $L=4$ and CRC-$8$
of \eqref{eq:CRC8gen}.

\begin{table}[t] 
  \caption{LLR-Based SC Decoder vs. SCL Decoder Synthesis Results}
  \label{tab:scvsscl} 
  \centering 
  \begin{tabular}{r|c|c||c|c}
    & \multirow{2}{*}{SC} & CA-SCLD & \multirow{2}{*}{SC} & CA-SCLD \\
    & & $L=2$, CRC-$4$ & & $L=4$, CRC-$8$\\
    \hline
    $N$ & $2048$ & $1024$ & $4096$ & $1024$ \\
    \hline
    Freq. (MHz) 		
    & $870$		& $847$ 
    & $806$		& $794$\\ 
    Lat. (Cyc./bit)	
    & $2.05$	& $2.54$
    & $2.06$	& $2.59$\\
    Lat. (Cyc.)		
    & $4192$	& $2596$
    & $8448$	& $2654$\\    
    T/P (Mbps)		
    & $425$		& $334$ 
    & $391$		& $306$\\
    Area (mm$^2$) 		
    & $0.78$	& $0.88$
    & $1.51$	& $1.78$\\
  \end{tabular}
\end{table}

As mentioned earlier, our SCL decoder architecture is based on the SC decoder 
of~\cite{Leroux13}. In Table~\ref{tab:scvsscl} we present the synthesis results
for the SC decoder of~\cite{Leroux13} at block lengths $N=2048$ and $N=4096$ and
compare them with that of our LLR-based SCL decoder, when using the same TSMC
$90$nm technology and identical operating conditions.  For all decoders, we use
$P = 64$ PEs per path and $Q=6$ bits for the quantization of the LLRs.

First, we see that the SCL decoders occupy an approximately $15\%$ larger area
than their SC decoder counterparts. This may seem surprising, as it can be
verified that an SC decoder for a code of length $LN$ requires more memory (LLR
and partial sum) than the memory (LLR, partial sum, and path) required by an SCL
decoder with list size $L$ for a code of length $N$, and we know that the memory
occupies the largest fraction of both decoders.  This discrepancy is due to the
fact that the copying mechanism for the partial sum memory and the path memory
still uses $L \times L$ crossbars, which occupy significant area.  It is an
interesting open problem to develop an architecture that eliminates the need for
these crossbars.

Moreover, we observe that both SC decoders can achieve a slightly higher
operating frequency than their corresponding SCL decoders, although the
difference is less than $3\%$.  However, the per-bit latency of the SC decoders
is about $20\%$ smaller than that of the SCL decoders, due to the sorting step
involved in SCL decoding. The smaller per-bit latency of the SC decoders
combined with their slightly higher operating frequency, make the SC decoders
have an almost $27\%$ higher throughput than their corresponding SCL decoders.

However, from Table~\ref{tab:scvsscl} we see that the SCL decoders have a
significantly lower per-codeword latency. More specifically, the SCL decoder
with $N = 1024$ and $L=2$ has a $38$\% lower per-codeword latency than the SC
decoder with $N = 2048$, and the SCL decoder with $N = 1024$ and $L = 4$ has a
$68$\% lower per-codeword latency than the SC decoder with $N = 4096$. Thus, for
a fixed FER, our LLR-based SCL decoders provide a solution of reducing
the per-codeword latency at a small cost in terms of area, rendering them more
suitable for low-latency applications than their corresponding SC decoders.
\subsection{Simplified SC and SCL Decoders}
There has been significant work done to reduce the latency of SC
decoders~\cite{Alam11,Zhang12,Sarkis13,Zhang14} by pruning the decoding graph,
resulting in \emph{simplified SC} (SSC) decoders.  The SC decoder architecture
of~\cite{Leroux13}, used in our comparison above, does not employ any of these
techniques. Since our SCL decoder uses $L$ SC decoders, it seems evident that
any architectural and algorithmic improvements made to the SC decoder itself
will be beneficial to the LLR-based SCL decoder as well. However, the family of
SSC decoders does not seem to be directly applicable to our LLR-based SCL
decoder. This happens because, in order to keep the path metric updated, we need
to calculate the LLRs even for the frozen bits. As discussed in
Section~\ref{sec:theory}, it is exactly these LLRs that lead to the improved
performance of the SCL decoder with respect to the SC decoder.  However,
alternative and promising pruning approaches which have been recently introduced
in the context of LL-based SCL decoding~\cite{Yuan14,Xion15}, are fully
applicable to LLR-based SCL decoding.
\section{Conclusion}\label{sec:conclusion}
In this work, we introduced an LLR-based path metric for SCL decoding of polar
codes, which enables the implementation of a numerically stable LLR-based SCL
decoder.  Moreover, we showed that we can simplify the sorting task of the SCL
decoder by using a pruned radix-$2L$ sorter which exploits the properties of the
LLR-based path metric. The LLR-based path metric is not specific to SCL decoding
and can be applied to any other tree-search based decoder (e.g., stack SC
decoding \cite{Chen13}).

We implemented a hardware architecture for an LLR-based SCL decoder and we
presented synthesis results for various list sizes.  Our synthesis results
clearly show that our LLR-based SCL decoder has a significantly higher
throughput \emph{and} lower area than all existing decoders in the literature,
leading to a substantial increase in hardware efficiency of up to $137\%$.

Finally, we showed that adding the CRC unit to the decoder and using CA-SCLD is
an easy way of increasing the hardware efficiency of our SCL decoder at a given
block-error probability as the list size can be decreased. Specifically, our
CA-SCLD at list size $L=2$ has somewhat lower block-error probability \emph{and}
more than five times better hardware efficiency than our standard SCLD at list
size $L=8$. 

\ifCLASSOPTIONcaptionsoff
  \newpage
\fi



%
\bibliographystyle{IEEEtran}
\bibliography{IEEEfull,refs}

%

%
%
%




\end{document}